\documentclass[reqno,11pt]{article}
\pdfoutput=1
\usepackage{jheppub}
\graphicspath{ {./graphs1/} }
\usepackage{epsfig}
\usepackage{amssymb}
\usepackage{verbatim}
\usepackage{amsmath}
\usepackage{hyperref}
\usepackage{dsfont}
\usepackage{slashed}
\usepackage[dvipsnames]{xcolor}
\usepackage{epstopdf}
\usepackage{graphicx}
\def\o{\omega}

\def\l{\lambda}

\def\L{\Lambda}

\def\s{\sigma}

\def\a{\alpha}
\def\b{\beta}

\newcommand{\be}{\begin{eqnarray}}
\newcommand{\en}{\end{eqnarray}}

\newcommand{\bc}{\begin{center}}
\newcommand{\ec}{\end{center}}

\title{Quadratic Feynman Loop Integrands From Massless Scattering Equations}
\author{Humberto Gomez}

\affiliation{Facultad de Ciencias Basicas,  Universidad Santiago de Cali,\\
Calle 5 $N^\circ$  62-00 Barrio Pampalinda, Cali, Valle, Colombia}

\emailAdd{humgomzu@gmail.com }

\abstract{Recently the Cachazo-He-Yuan (CHY) approach has been extended to loop level, but 
the  resulting loop integrand has propagators that are linear in the loop momentum unlike Feynman's. 
In this note we present a new technique that directly produces quadratic propagators identical to Feynman's from the CHY approach.
This paper focuses on $\Phi^3$ theory but extensions to others theories are briefly discussed. In addition, our proposal has an interesting geometric meaning, we can interpret this new formula as a unitary cut on a higher genus Riemann surface.
}

\begin{document}

\maketitle

%%%%%%% Introduction & Summary of the Results %%%%%%%
%%%%%%%%%%%%%%%%%%%%%%%%%%%%%%%%%%%%%%%%%%%%%%%%%%%%%

%%%%%%%%%%%%%%%%%%%%%%%%%%%%%%%%%%
\section{Introduction}\label{sect:intro}
%%%%%%%%%%%%%%%%%%%%%%%%%%%%%%%%%%

Since the remarkable work of Witten \cite{Witten:2003nn} on the ${\cal N}=4$ super Yang--Mills theory, the
on-shell methods for the computation of scattering amplitudes have been deeply studied during the last years. In particular, the Cachazo--He--Yuan (CHY)  approach \cite{Cachazo:2013gna, Cachazo:2013hca,Cachazo:2013iaa,Cachazo:2013iea}, which is applicable in arbitrary dimension,  is an outstanding method because it can be applied for a large family of interesting theories including scalars, gauge bosons, gravitons and mixing interactions among them \cite{Cachazo:2014xea, Cachazo:2014nsa, Cachazo:2016njl}. The original proposal  is to write the tree-level S-matrix in terms of a contour integral  localized over solutions of the so-called {\it scattering equations} \cite{Cachazo:2013gna} on the moduli space of $n$-punctured Riemann spheres. Other approaches that use the same moduli space include the Witten--RSV \cite{Witten:2003nn,Roiban:2004yf}, Cachazo--Geyer \cite{Cachazo:2012da}, and Cachazo--Skinner \cite{Cachazo:2012kg} constructions, but are special to four dimensions.

The CHY formalism has already been verified to reproduce well-known results, such as the soft limits of various theories \cite{Cachazo:2013hca}, the Kawai--Lewellen--Tye relations \cite{Kawai:1985xq} between gauge and gravity amplitudes \cite{Cachazo:2013gna}, as well as the correct Britto--Cachazo--Feng--Witten \cite{Britto:2005fq} recursion relations in Yang--Mills and bi-adjoint $\Phi^3$ theories \cite{Dolan:2013isa}.

Nevertheless, a direct evaluation of the CHY-integrals for higher order poles is not a simple task. So, many methods have been developed during the last year to deal with them. These attempts include the study of solutions at particular kinematics and/or dimensions \cite{Cachazo:2013iea, Kalousios:2013eca, Lam:2014tga, Cachazo:2013iaa, Cachazo:2016sdc, He:2016vfi, Cachazo:2015nwa, Cachazo:2016ror}, encoding the solutions to the scattering equations in terms of linear transformations \cite{Kalousios:2015fya,Dolan:2014ega, Huang:2015yka, Cardona:2015ouc, Cardona:2015eba, Dolan:2015iln, Sogaard:2015dba, Bosma:2016ttj, Zlotnikov:2016wtk,Mafra:2016ltu}, or the formulation of integration rules in terms of the polar structures \cite{Baadsgaard:2015ifa,Baadsgaard:2015voa,Huang:2016zzb,Cardona:2016gon}.  In particular, the author in \cite{Gomez:2016bmv} gave  an independent proposal by generalizing the double-cover formulation, the so-called {\it $\Lambda$-algorithm}, which we are going to use in this paper.

The CHY formalism has been generalized to loop level. Using the ambitwistor and pure spinor ambitwistor string \cite{Mason:2013sva, Berkovits:2013xba, Gomez:2013wza}, a proposal was made in \cite{Geyer:2015bja, Geyer:2015jch, Adamo:2015hoa}. Parallelly, in \cite{Cachazo:2015aol,He:2015yua, Baadsgaard:2015hia}, another prescription was developed by performing a forward limit on the scattering equations for massive particles formulated previously in \cite{Naculich:2014naa, Dolan:2013isa}. On the other hand, the author {\it et al}, following the ideas presented in \cite{Gomez:2016bmv}, obtained an alternative formulation at one-loop by embedding the torus in a $\mathbb{CP}^2$ through an elliptic curve \cite{Cardona:2016bpi} and used it to reproduce the $\Phi^3$ theory at one loop \cite{Cardona:2016wcr}. Recently,  in \cite{Chen:2016fgi,Chen:2017edo} a differential operators on the moduli space were created  to compute CHY-integrals at one-loop.  In addition, extensions at two-loop are being studied and some important results have been found in \cite{Geyer:2016wjx, Feng:2016nrf,Gomez:2016cqb}.

\begin{figure}[!h]
  \centering
         \includegraphics[width=1.4in]{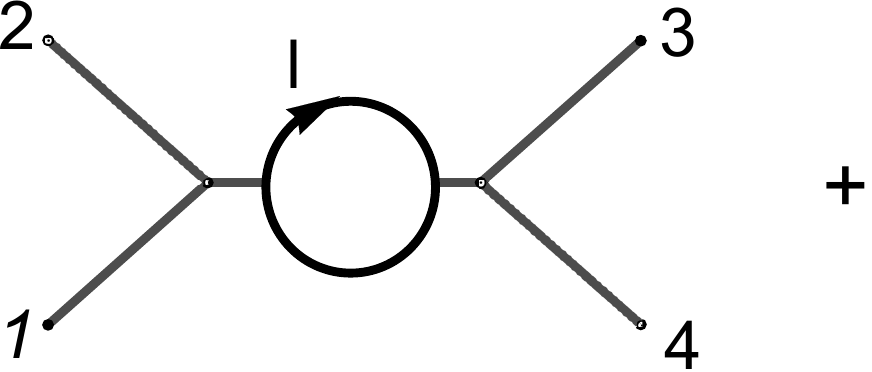}     \quad                               
         \includegraphics[width=1.1in]{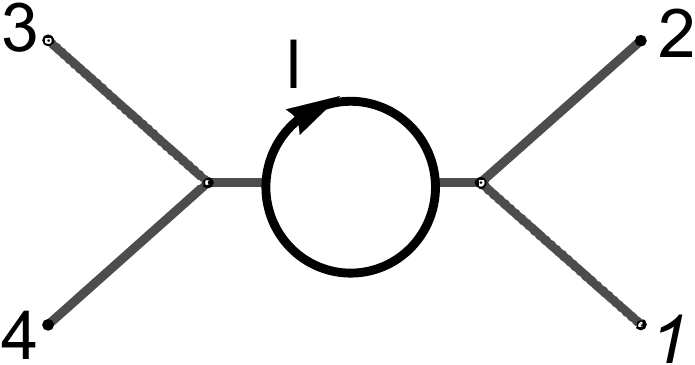}
  \caption{$\Phi^3$ symmetrized Feynman diagram.}\label{intro_ex}
\end{figure}

However, all results obtained at loop level from the CHY approach do not match in an exact way with the Feynman diagrams.   In order to establish an equivalence among the CHY approach and the Feynman integrads at loop level,  it is needed to use the partial fraction identity (${\rm p.f}$)
\begin{equation*}
{1\over \prod_{i=1}^n D_i}=\sum_{i=1}^{n}{1\over D_i\prod_{j\neq i}(D_j-D_i)}\, 
\end{equation*}
and after shifting (${\rm S}$) the internal loop momentum. This is not a trivial process and its real matching is still analized \cite{Huang:2015cwh, Baadsgaard:2015twa,Feng:2016msc}.  For example, let us consider the symmetrized $\Phi^3$ Feynman diagram in figure \ref{intro_ex}.
%%%%%%%%%%%%%%%%%%%%%%%%%%%%%
\noindent
\\
The expression for this amplitude is given by\footnote{Since in this paper we care just for the Feynman integrands, then the dimension of the space-time and the convergence of the loop integrals will not be an issue to be discussed here.}  
\begin{equation*}
{\cal A}^{\rm FEY}=\int d^D\ell\,\,{\cal I}^{\rm FEY}=  \int \frac{d^D\ell}{s_{12}s_{34}} \left[ \frac{1}{ \ell^2\,(\ell + k_1+k_2)^2} +\frac{1}{ \ell^2\,(\ell - k_1-k_2)^2} \right].
\end{equation*}
Now, from the CHY approach at one-loop\footnote{We are going to do a quick review in section \ref{sectionone}.} \cite{Cardona:2016bpi,Cardona:2016wcr},   the result for the same process is 
\begin{equation*}
{\cal A}^{\rm CHY}=\int d^D\ell\,\,{\cal I}^{\rm CHY}= 2 \int \frac{d^D\ell}{s_{12}s_{34}} \left[ \frac{1}{\ell^2\,\left[(\ell + k_1+k_2)^2-\ell^2\right]} +\frac{1}{ \ell^2\left[ (\ell - k_1-k_2)^2-\ell^2\right]} \right],
\end{equation*}
which is referred to as the \emph{Q-cut representation} \cite{Baadsgaard:2015twa}. 
In order to obtain an equivalence at the integrand level, we apply the partial fraction identity in ${\cal A}^{\rm FEY}$, so it becomes
\begin{align*}
{\cal A}^{\rm FEY}\Big|_{\rm p.f}= &\int \frac{d^D\ell}{s_{12}s_{34}} \left[ \frac{1}{\ell^2\,\left[(\ell + k_1+k_2)^2-\ell^2\right]} +\frac{1}{(\ell + k_1+k_2)^2\,\left[\ell^2-(\ell + k_1+k_2)^2\right]}
\right.\\
&\qquad\qquad\left.
+\frac{1}{ \ell^2\left[ (\ell - k_1-k_2)^2-\ell^2\right]} 
+\frac{1}{ (\ell - k_1-k_2)^2 \left[ \ell^2- (\ell - k_1-k_2)^2\right]} 
\right].
\end{align*}
At last, by performing the shift transformations, $\ell^\prime = \ell +k_1+k_2$ and $\ell^\prime = \ell -k_1-k_2$, in the second and fourth term on ${\cal A}^{\rm FEY}\Big|_{\rm p.f}$, respectively, it is simple to check how the amplitude ${\cal A}^{\rm FEY}$  becomes  ${\cal A}^{\rm CHY}$.

In this work we give, for first time, a proposal which is able to reproduce the physical quadratic Feynman integrand\footnote{By physical quadratic Feynman integrand we mean the integrand as computed with standard Feynman diagrams.} at one-loop from the CHY approach, i.e. it is not necessary to use the partial fraction identity nor shifting the loop momentum. Although in this paper we are just focused to the $\Phi^3$ theory, we are working to extend our ideas to other theories and also at two-loop \cite{wp}. Our formula, which is motived from two-loop prescription given in \cite{Gomez:2016cqb},  is simple and it includes into itself the results found in \cite{Cardona:2016bpi,Cardona:2016wcr}, as it is shown in section \ref{examples} and argued in section \ref{conclusion}. 

One of the main virtues of this new proposal is that the CHY integrals are localized on the original scattering equations given in \cite{Cachazo:2013gna, Cachazo:2013hca,Cachazo:2013iaa}. Let us be more explicit, if we wish  to obtain the quadratic Feynman integrand of a scattering of $n$ massless particles at one-loop using this new CHY proposal, then we must solve the scattering equations of $(n+4)$ on-shell particles, i.e.
\begin{equation*}
E_A = \sum_{B=1 \atop B\neq A}^{n+4}\frac{k_A\cdot k_B}{\s_{AB}} =0, \qquad k_A^2=0, \qquad A=1,\ldots, n+4. 
\end{equation*}
This means that all methods developed to compute CHY integrals of massless particles can be used without any restriction, in particular the $\L-$algorithm \cite{Gomez:2016bmv}, which will be used in section \ref{examples}.

\begin{figure}[!h]
%   Requires \usepackage{graphicx}
\centering
\raisebox{2.0\height}{\includegraphics[scale=.4]{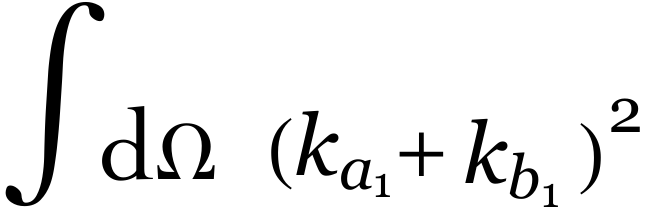}}\quad
                                  \includegraphics[scale=0.4]{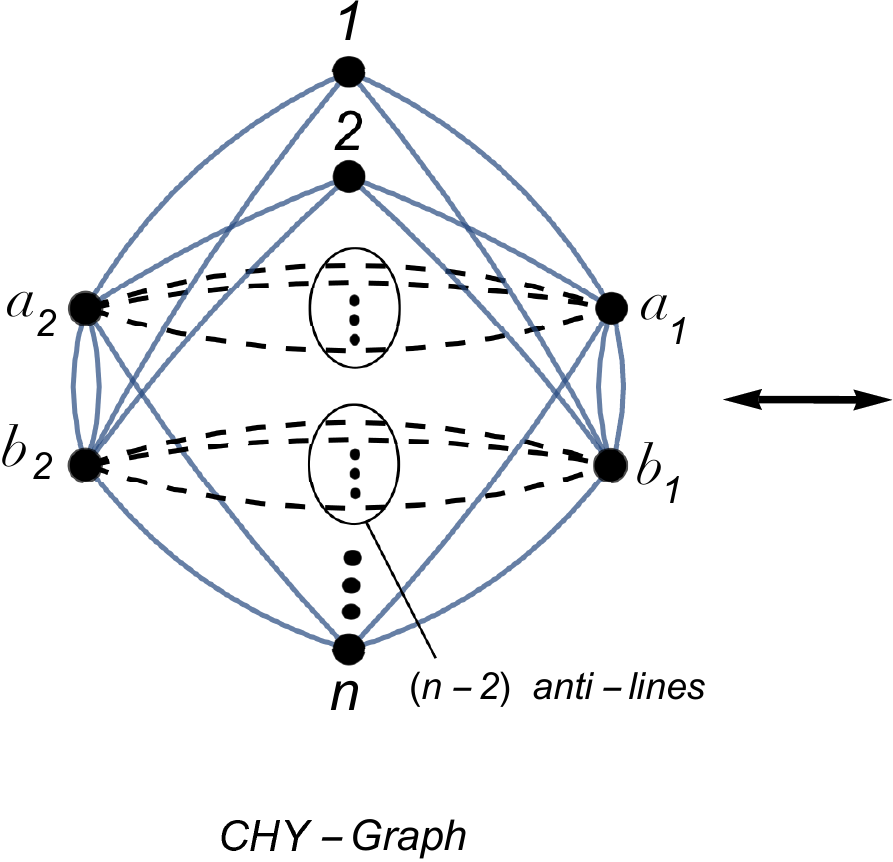}
\raisebox{1.3\height}{\includegraphics[scale=.13]{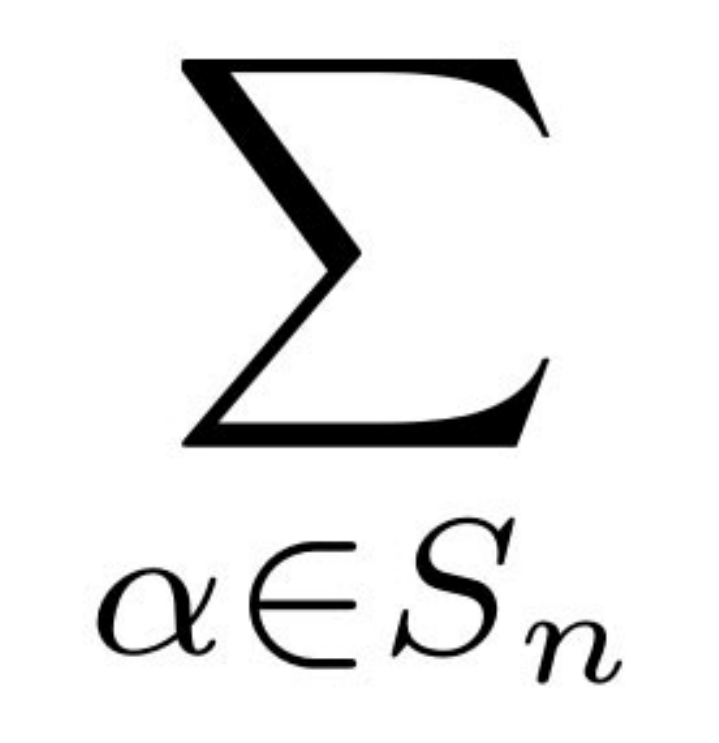}}
 \raisebox{0.1\height}{\includegraphics[scale=0.5]{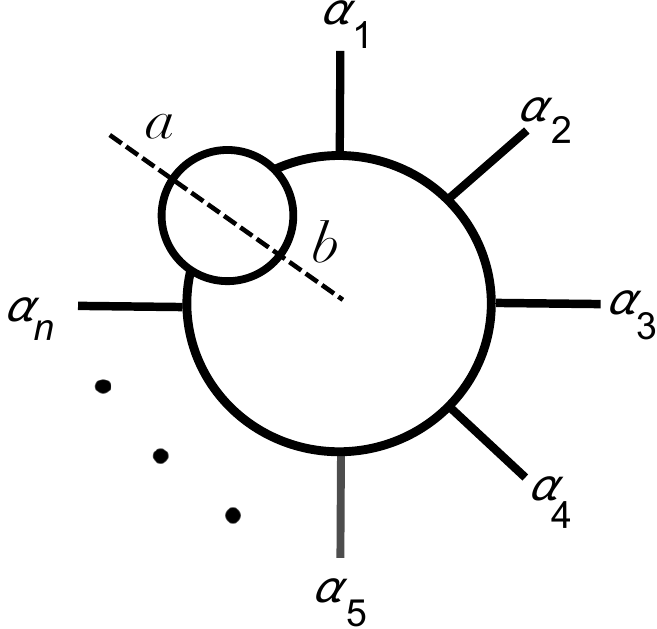}}
  \caption{Schematic representation for the new CHY proposal.}\label{intro_2}
\end{figure}

\begin{figure}[!h]
%   Requires \usepackage{graphicx}
\centering
\raisebox{1.2\height}{\includegraphics[scale=.4]{int_dO-eps-converted-to.pdf}}\quad
                                  \includegraphics[scale=0.4]{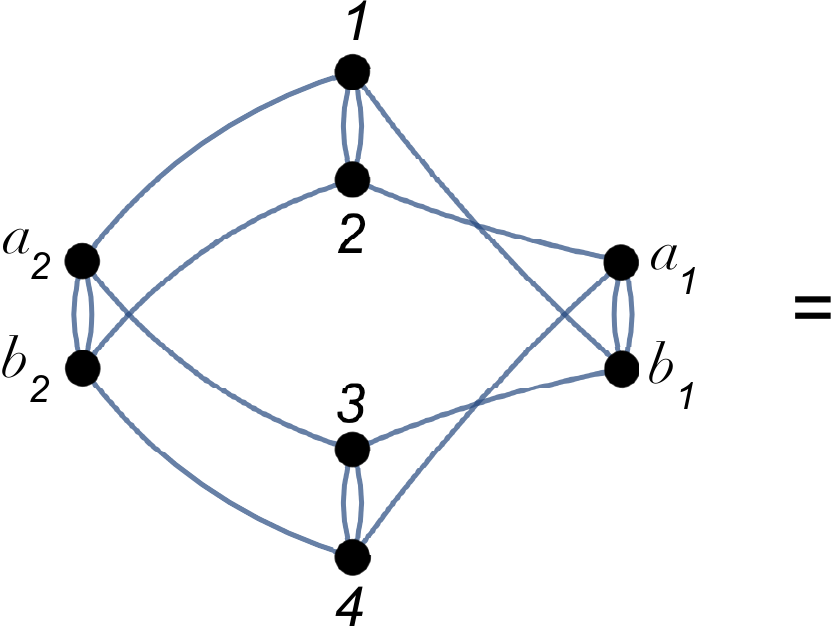}\quad
  \raisebox{0.3\height}{\includegraphics[width=1.4in]{intro_1-eps-converted-to.pdf}}     \quad                               
     \raisebox{0.3\height}{\includegraphics[width=1.1in]{intro_2-eps-converted-to.pdf}}
  \caption{Equality among CHY and Feynman integrands.}\label{intro_3}
\end{figure}

Finally, one can summarize saying we have found a novel form to write the scattering amplitudes at one-loop from an unitary cut on a Riemann surface of genus two. Schematically our result is represented in figure \ref{intro_2}, 
%%%%%%%%%%%%%%%%%%%%%%%%%%%%%
\noindent
\\
where the measure $d\Omega$ is defined as 
\begin{align*}
d\Omega :&= d^D(k_{a_1}+k_{b_1})\, \delta^{(D)}(k_{a_1}+k_{b_1} - \ell)\, d^{D}k_ {a_2}\, d^{D}k_{ b_2}\,\delta^{(D)} (k_{a_2}+k_{a_1})\,\,\delta^{(D)} (k_{b_2}+k_{b_1}).
\end{align*}
and the Dirac delta functions in $d\Omega$ guarantees the unitary cut. For instance, let us consider the Feynman integrand given in the above example in the amplitude ${\cal A}^{\rm FEY}$, figure \ref{intro_ex}. So, from this new proposal the CHY integrand represented in figure \ref{intro_3} is able to reproduce the same ${\cal I}^{\rm FEY}$ integrand.
%%%%%%%%%%%%%%%%%%%%%
%%%%%%%%%%%%%%%%%%%%%%%%%%%%%
\noindent
\\
This new representation is going to be explained in detail in section \ref{sectwo} and  \ref{conclusion}.

%%%%%%%%%%%%%%%%%%%%%
\subsection*{Outline}
%%%%%%%%%%%%%%%%%%%%%

This paper is organized in the following way. In section \ref{sectionone} we present a general review of the  CHY approach at one-loop for the $\Phi^3$ theory. As we emphasize, the equivalence with the Feynman diagrams is obtained after using the partial fraction identity and by shifting the loop momentum.  In section \ref{sectwo} we propose a new formula to obtain  quadratic Feynman integrands from the CHY approach. Notice that the new CHY-integrand proposed in \eqref{CHY-I} is totally similar to the one already known given in \eqref{CHY_oneL}, with the big difference that there are two more particles. In section \ref{examples} we give some examples in order to illustrate our new formula proposed in section \ref{sectwo}.  The computations are presented in detail, so in each example one can see as the old one-loop prescription, given in section \ref{sectionone}, contributes to the final answer. To end, in section \ref{conclusion}  we present some conclusions and make clear our motivation to propose the formula given in section \ref{sectwo}. In addition, a few perspectives are put in context.

Before beginning with the review at one-loop, we define the notation that is going to be used in the paper.

%%%%%%%%%%%%%%%%%%%%%
\subsection*{Notation}
%%%%%%%%%%%%%%%%%%%%%

In order to have a graphical description for the CHY integrands on a Riemann sphere (CHY-graphs), it is convenient to represent the factor ${1 \over \s_{ab}}$ as a {\it line} and the factor $\s_{ab}$ as a dashed line that we call the {\it anti-line}:
\begin{align}
	&{1\over \s_{ab}}\,\leftrightarrow\,a ~\overline{~~~~~~~~~~~~~}~ b ~~~{\rm (line)},  \\
	& \s_{ab}~\leftrightarrow\, a\,- \, - \,  - \, -\, b ~~~     {\rm  (anti {\rm -} line)}.
\end{align}
%In this way, CHY integrands have graphical description as CHY graphs

\begin{figure}[!h]
	%Requires \usepackage{graphicx}
	\centering
	\includegraphics[width=5.0in]{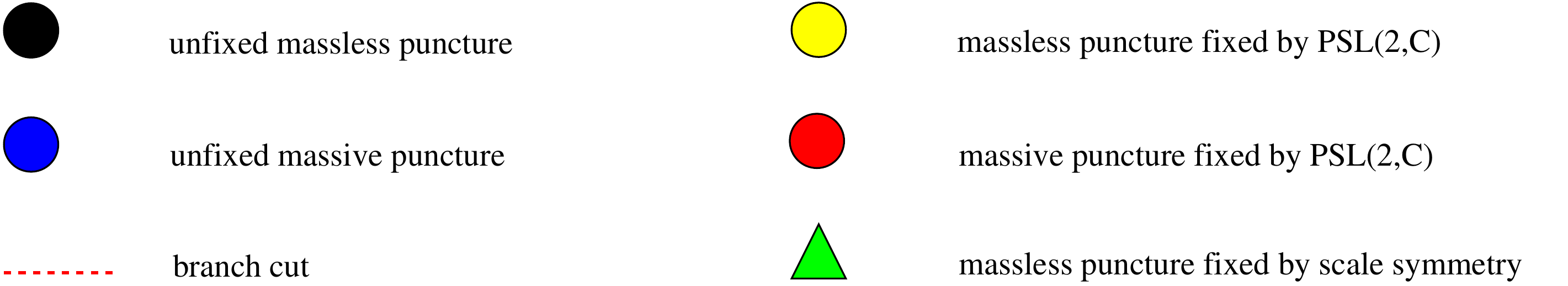}
	\caption{Color code in CHY-graphs.}\label{color_cod}
\end{figure}

Additionally, since we often use the CHY-graphs and the $\L-$algorithm\footnote{It is useful to recall that the $\L-$algorithm fixes four punctures, three of them by the ${\rm PSL}(2,\mathbb{C})$ symmetry and the last one by the scale invariance.} \cite{Gomez:2016bmv}, it is useful to introduce the color code given in figure \ref{color_cod}
for a  mnemonic understanding.
\noindent

Finally, we introduce the following notation
\begin{align}\label{notations}
&
k_{\{ a_1,\ldots, a_m\}}= [a_1,\ldots, a_m]:=\sum_{i=1}^m k_{a_i}, \quad
s_{a_1\ldots a_m}:=k_{\{ a_1,\ldots, a_m\}}^2, 
\quad k_{a_1\ldots a_m}:=\sum_{a_i<a_j}^m k_{a_i}\cdot k_{a_j}, \nonumber \\
& \s_{ij} := \s_i - \s_j, 
\quad  (i_1,i_2,\ldots, i_p):= \s_{i_1i_2} \cdots \s_{i_{p-1}i_p}\s_{i_p i_1},
\quad \o_{i:j}^{a:b} := \frac{\s_{ab}}{\s_{i a}\,\, \s_{j b}}, \\
& B(A,B,C,D) : = \frac{1}{k_A\cdot k_C + \frac{1}{2}k_A^2} +\frac{1}{k_A\cdot k_D + \frac{1}{2}k_A^2},\nonumber
\end{align}
where $A,B,C$ and $D$ are sets, for example $A=\{ a_1,a_2,\ldots a_p\}$, and $k_A=k_{\{ a_1,a_2,\ldots a_p\}}$. Note that $\o_{i:j}^{a:b}$ is the generalization of the $(1,0)-$forms used in \cite{Gomez:2016cqb} to write the CHY integrands at two-loop.

%%%%%%%%%%%%%%%%%%%%%%%%%%%%%%%%%%%%%%%%%%%%%%%%%%%%%%
%%%%%%%%%%%%%%%%%%%%%%%%%%%%%%%%%%%%%%%%%%%%%%%%%%%%%%
\section{Simple Review of CHY $\Phi^3$ Integrand at One-Loop }\label{sectionone}
%%%%%%%%%%%%%%%%%%%%%%%%%%%%%%%%%%%%%%%%%%%%%%%%%%%%%%
%%%%%%%%%%%%%%%%%%%%%%%%%%%%%%%%%%%%%%%%%%%%%%%%%%%%%%

In this section we present a simple and quick review about CHY integrands at one-loop. This is going to be very useful in order to compare and understand our new results.

\begin{figure}[!h]
  \centering
   \raisebox{.65\height}{\includegraphics[scale=.2]{Sum_Sp-eps-converted-to.pdf}}
         \includegraphics[width=1.9in]{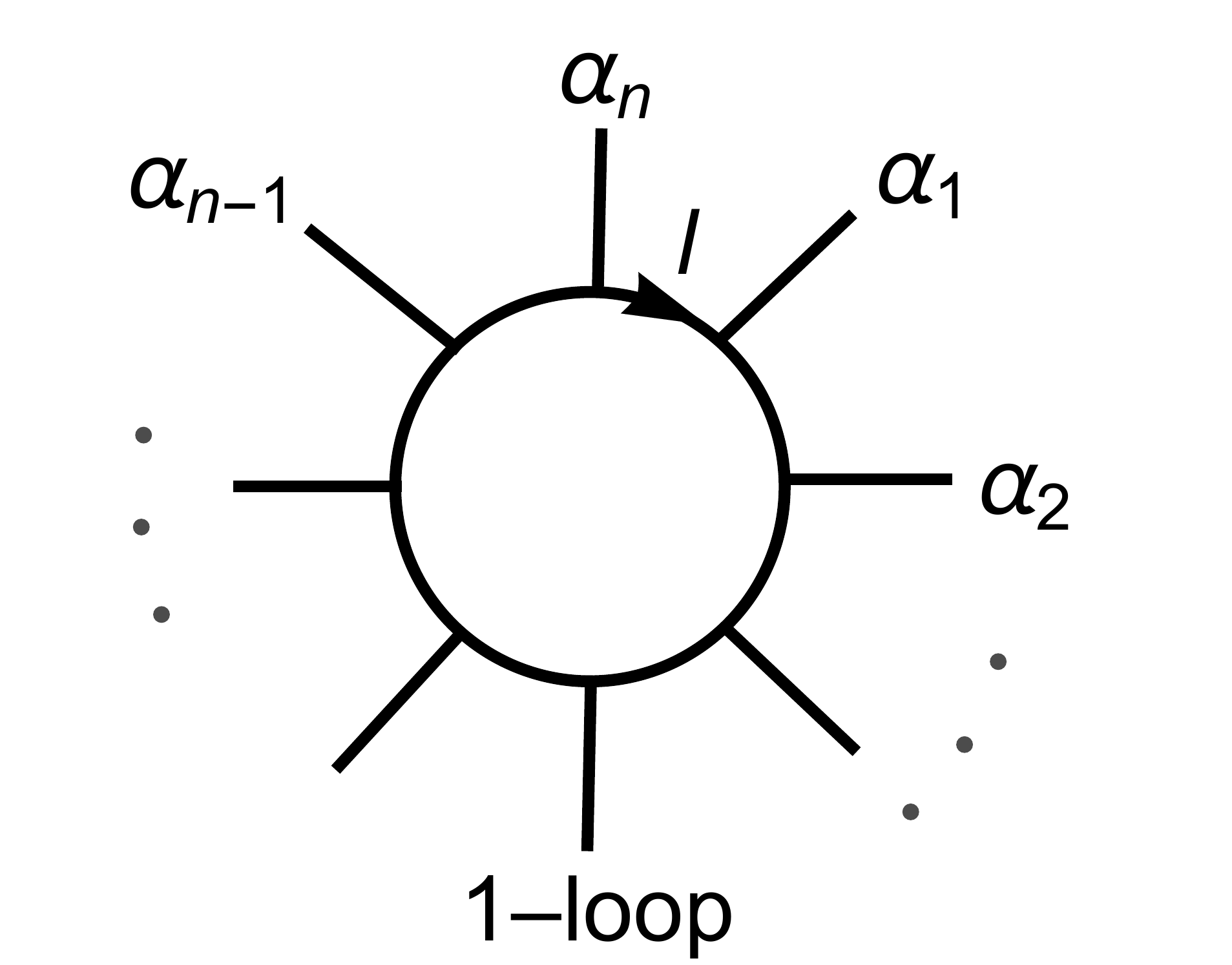}                                               
         \includegraphics[width=1.5in]{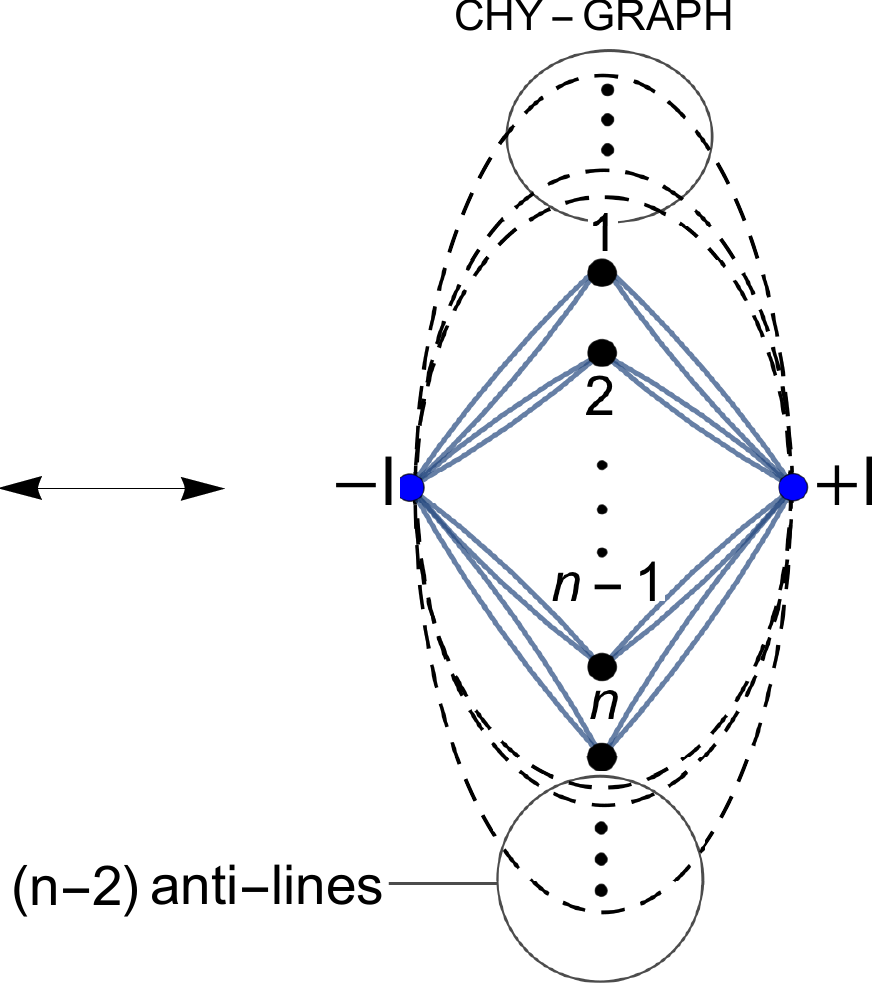}\\
  \caption{Correspondence between the $\Phi^3$ Feynman diagrams ($n$-gon-symmetrized)  and the CHY graphs, up to partial fraction identity and $\frac{1}{\ell^2}$ factor. $S_n$ is the permutation group and $\a_i : = \a(i)$.}\label{CHY-pgon-sym}
\end{figure}

As it was shown  in \cite{Cardona:2016bpi,Cardona:2016wcr,Geyer:2015bja,Geyer:2015jch,Gomez:2016cqb,He:2015yua,Chen:2016fgi}, the CHY integrand at one loop of the  symmetrized $n$-gon can be written as
\begin{align}\label{CHY_oneL}
{\bf I}^{\rm n-gon}_{\rm sym}&=  \frac{1}{(\ell^+, \ell^- )^2} \left(\o_{1:2}^{\ell^+:\ell^-}\,\o^{\ell^+:\ell^-}_{2:3}\, \cdots \o^{\ell^+:\ell^-}_{n:1}\right)\times \left(\o^{\ell^+:\ell^-}_{1:n}\,\o^{\ell^+:\ell^-}_{n-1:n-2}\, \cdots \o^{\ell^+:\ell^-}_{2:1}\right),
\end{align}
The CHY-graph representation for  $ {\bf I}^{\rm n-gon}_{\rm sym}$ is given on the right side  in figure \ref{CHY-pgon-sym}.
%%%%%%%%%%%%%%%%%%%
%%%%%%%%%%%%%%%%%%%%%%%%%%%%%
\noindent

By integrating ${\bf  I}^{\rm n-gon}_{\rm sym}$ over the moduli space, i.e.
\begin{equation}
{\cal I}^{\rm n-gon-CHY}_{\rm sym}:={1\over \ell^2}\int d\mu^{\rm 1-loop}\,\,\,  I^{\rm n-gon}_{\rm sym},
\end{equation}
with 
\begin{align}
d\mu^{\rm1-loop} :=&\frac{1}{{\rm Vol}\,\,({\rm PSL}(2,\mathbb{C}))}\times \frac{d\s_{\ell^+}}{E^{\rm 1-loop}_{\ell^+}}\times\frac{d\s_{\ell^-}}{E^{\rm 1-loop}_{\ell^-}}\times \prod_{a=1}^n \frac{d\s_{a}}{E^{\rm 1-loop}_{a}},\nonumber\\
&   ^{\underrightarrow{\,\quad{\rm fixing\,\, PSL}(2,\mathbb{C})\,\quad }}\,\,\,\prod_{i=1}^{n-1} \frac{d\s_{i}}{E^{\rm 1-loop}_{i}}\times (\s_{\ell^+\ell^-}\,\s_{\ell^- n}\, \s_{n \ell^+}  )^2
\end{align}
and
\begin{align}
&E_a^{\rm 1-loop}:= \sum^n_{b=1\atop b\neq a}\frac{k_a\cdot k_b}{\s_{ab}} +  \frac{k_a\cdot \ell^+}{\s_{a\ell^+}} + \frac{k_a\cdot \ell^-}{\s_{a\ell^-}},\\
& E_{\ell ^{\pm}}^{\rm 1-loop}:= \sum^{n}_{b=1}\frac{\ell^{\pm}\cdot k_b}{\s_{\ell^{\pm}b}}, 
\qquad  (\ell^+)^\mu=-(\ell^-)^\mu := \ell^\mu, \quad \ell^2\neq 0. 
\end{align}
where, without loss of generality,  we have fixed $\{\s_{\ell^+},\s_{\ell^-},\s_{n} \}$
and  $\{ E^{\rm 1-loop}_{\ell^+},E^{\rm 1-loop}_{\ell^-},E^{\rm 1-loop}_{n} \}$,
it was argued in  \cite{Geyer:2015bja,Geyer:2015jch,He:2015yua} that ${\cal I}^{\rm n-gon-CHY}_{\rm sym}$ becomes
\begin{equation}
{\cal I}^{\rm n-gon-CHY}_{\rm sym} = {1\over \ell^2}\sum_{\a\in S_n}\frac{1}{(\ell\cdot k_{\a_1}) (\ell\cdot (k_{\a_1}+k_{\a_2})+k_{\a_1}\cdot k_{\a_2}) \ldots (-\ell\cdot k_{\a_n})},
\end{equation}
where $\a_i:=\a(i)$ and $S_n$ is the permutation group of $n-$elements, $S_n:= {\rm Permutations}\,\{1,2,\ldots, n \}$.

On the other hand, the Feynman integrand for the symmetrized diagram on the left side in figure \ref{CHY-pgon-sym} is given by
\begin{equation}\label{Fey_oneloop}
{\cal I}^{\rm n-gon-FEY}_{\rm sym} =\frac{1}{\ell^2} \sum_{\a\in S_n}\frac{1}{
(\ell+ k_{\a_1})^2 (\ell+k_{\a_1}+k_{\a_2})^2\ldots  (\ell- k_{\a_n})^2 }.
\end{equation}

The equivalence among ${\cal I}^{\rm n-gon-CHY}_{\rm sym}$ and  $\,{\cal I}^{\rm n-gon-FEY}_{\rm sym}$ is  established after using the {\bf partial fraction identity (${\rm p.f}$)}
\begin{equation}\label{partialfrac}
{1\over \prod_{i=1}^n D_i}=\sum_{i=1}^{n}{1\over D_i\prod_{j\neq i}(D_j-D_i)}\, 
\end{equation}
and by assuming  that the loop integral, $\int d^D\ell$, is invariant under {\bf shifting ($\rm S$)} of the loop momentum $\ell$, so\footnote{Here the factor $2^{-n+1}$ comes from the convention of using $k_a\cdot k_b$  instead of $2k_a\cdot k_b$ in the numerators of the scattering equations. In a general $l$-loop case, this factor is $2^{-(n+2l-3)}$ due to the ${\rm PSL}(2,\mathbb{C})$ symmetry of scattering equations and the number of puncture locations \cite{Gomez:2016bmv}. The number $2$ in the numerator comes from the $\mathbb{Z}_2$ symmetry $\ell \leftrightarrow -\ell$.} 
\begin{equation}\label{equi_oneloop}
{\cal I}^{\rm n-gon-FEY}_{\rm sym}\Big|_{\rm p.f \atop S}
= \frac{2}{2^{n-1}}\,\,
{\cal I}^{\rm n-gon-CHY}_{\rm sym}.
\end{equation}

%%%%%%%%%%%%%%%%%%%%%%%%%%%%%%%%%%%%%%%%%%%%
%%%%%%%%%%%%%%%%%%%%%%%%%%%%%%%%%%%%%%%%%%%
\section{Massless Scattering Equations and $\Phi^3$ Quadratic Feynman Loop Integrands}\label{sectwo}
%%%%%%%%%%%%%%%%%%%%%%%%%%%%%%%%%%%%%%%%%%%
%%%%%%%%%%%%%%%%%%%%%%%%%%%%%%%%%%%%%%%%%%%%

In this section we propose a new $\Phi^3$ CHY formula at one-loop, which is able to reproduce the one-loop quadratic Feynman integrand, i.e. it is not necessary to use the partial fraction identity given  in \eqref{partialfrac}.

The meaning and motivation to propose this new formula is going to be discussed in detail in section \ref{conclusion}.

%%%%%%%%%%%%%%%%%%%%%%%%%%%%
\subsection{A New CHY Proposal at One-Loop for $\Phi^3$}
%%%%%%%%%%%%%%%%%%%%%%%%%%%%

\begin{figure}[!h]
  \centering
 \includegraphics[scale=0.6]{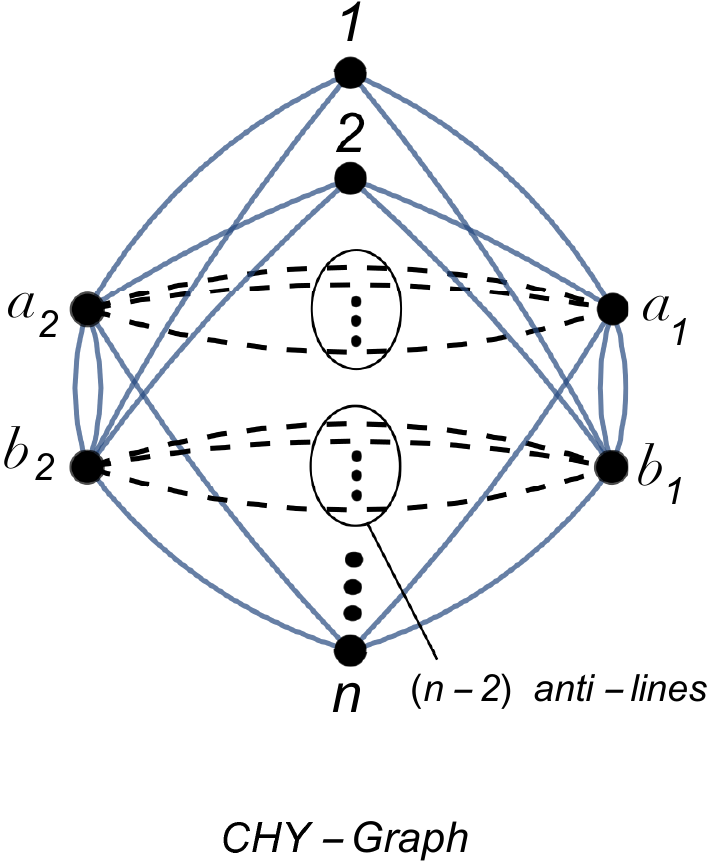}
  \caption{Graph representation for the CHY integrand, ${\bf I}^{\rm CHY}$, given in \eqref{CHY-I}.}\label{CHY_prescription}
\end{figure}

Let us consider the CHY integrand 
\begin{equation}\label{CHY-I}
\mathbf{I}^{\rm CHY} = 
\frac{1}{(a_1,b_1,b_2,a_2)^2}\,\,(\o^{a_1:a_2}_{1:2}\,\o^{a_1:a_2}_{2:3}\cdots \o^{a_1:a_2}_{n:1})\times (\o^{b_1:b_2}_{1:n}\,\o^{b_1:b_2}_{n-1:n-2}\cdots \o^{b_1:b_2}_{2:1}),
\end{equation}
which is represented graphically  in figure \ref{CHY_prescription}.
%%%%%%%%%%%%%%%%%%%
%%%%%%%%%%%%%%%%%%%%%%%%%%%%%
\noindent
\\

From the color code given in figure \ref{color_cod},  we are considering
all particles in figure \ref{CHY_prescription} are  massless, i.e.  $k_1^2=\cdots=k_n^2= k_{a_1}^2=k_{a_2}^2=k_{b_1}^2=k_{b_2}^2=0$. So, we should use the original scattering equations formulated  in \cite{Cachazo:2013hca,Cachazo:2013iaa,Cachazo:2013gna}. These equations are given by  the simple expressions
\begin{equation}
E_A = \sum_{B=1\atop B\ne A }^{n+4}\frac{k_A\cdot k_B}{\s_{AB}}=0,  \qquad A=1,2,\ldots, n+4,
\end{equation}
where we have identified
\begin{align}\label{identification}
\{ E_{n+1} , E_{n+2},E_{n+3},E_{n+4}\}&:=\{ E_{a_1} , E_{a_2},E_{b_1},E_{b_2}\}\nonumber,\\
\{ \s_{n+1} , \s_{n+2},\s_{n+3},\s_{n+4}\}&:=\{ \s_{a_1} , \s_{a_2},\s_{b_1},\s_{b_2}\},\\
\{ k_{n+1} , k_{n+2},k_{n+3},k_{n+4}\}&:=\{ k_{a_1} , k_{a_2},k_{b_1},k_{b_2}\}\nonumber,
\end{align}
and  the momentum conservation constraint, $\sum_{A=1}^{n+4}k_A=\sum_{i=1}^n k_i + k_{a_1}+ k_{a_2}+ k_{b_1}+ k_{b_2}=0$, is also satisfied.

Now, in order to obtain the $\Phi^3$ quadratic Feynman integrand at one-loop given in \eqref{Fey_oneloop} from the original CHY approach, we propose the following formula
\begin{equation}\label{prescription}
\mathfrak{I}^{\rm n-gon-CHY}_{\rm sym} : =\frac{1}{2^{n+1}}\int d\Omega\times(k_{a_1}+k_{b_1})^2\times  \int d\mu \,\,\,
\mathbf{I}^{\rm CHY},
\end{equation}
where $\mathbf{I}^{\rm CHY}$ is the CHY integrand represented in figure \ref{CHY_prescription} and given in \eqref{CHY-I},  $d\mu$ is the original CHY tree-level meausre \cite{Cachazo:2013hca}, namely
\begin{align}
d\mu:=& 
 \frac{1}{{\rm Vol}\,\,({\rm PSL}(2,\mathbb{C}))} \times \prod_{A=1}^{n+4}\frac{d\s_A}{E_A}\nonumber\\
 &   ^{\underrightarrow{\,\quad{\rm fixing\,\, PSL}(2,\mathbb{C})\,\quad }}\,\,\,
\frac{d\s_{a_1}}{E_{a_1}} \times
 \prod_{i=1}^{n} \frac{d\s_{i}}{E_{i}}\times (\s_{a_2b_1}\,\s_{b_1b_2}\, \s_{b_2a_2}  )^2
\end{align}
where, without loss of generality,  we have fixed $\{\s_{a_2},\s_{b_1},\s_{b_2} \}$
and  $\{ E_{a_2},E_{b_1},E_{b_2} \}$,
and $d\Omega$ is defined as
\begin{align}\label{dOmega}
d\Omega :&= d^D(k_{a_1}+k_{b_1})\, \delta^{(D)}(k_{a_1}+k_{b_1} - \ell)\, d^{D}k_ {a_2}\, d^{D}k_{ b_2}\,\delta^{(D)} (k_{a_2}+k_{a_1})\,\,\delta^{(D)} (k_{b_2}+k_{b_1}).
\end{align}

The Dirac delta functions, $\delta^{(D)} (k_{a_2}+k_{a_1})$ and $\delta^{(D)} (k_{b_2}+k_{b_1})$, are introduced to guarantee the forward limit\footnote{Note that we  can also impose the forward limit condition, $k_{a_2}=-k_{b_1}$ and $k_{b_2}=-k_{a_1}$, and the final answer will be the same.}, i.e. $k_{a_2}=-k_{a_1}$ and $k_{b_2}=-k_{b_1}$, and the last one, $\delta^{(D)}(k_{a_1}+k_{b_1} - \ell)$,  is given to identify $k_{a_1}+k_{b_1}$ with the 
the loop momentum $\ell$, i.e the off-shell momentum $\ell$ is represented as a couple of massless particles.

Using the $\L-$algorithm\footnote{Although all our computations have been performed by applying the $\L-$algorithm, any other algorithm can be used, such as ones given in \cite{Baadsgaard:2015voa, Cardona:2016gon, Cardona:2015eba,Cardona:2015ouc, Cachazo:2015nwa, Huang:2016zzb, Huang:2015yka,Kalousios:2015fya}.  
} developed by the author in \cite{Gomez:2016bmv}, we have verified, up to eight points, that in fact, the prescription proposed in \eqref{prescription} reproduces the quadratic Feynman integrand  given in \eqref{Fey_oneloop}. So, we conjecture 
\begin{align}
\mathfrak{I}^{\rm n-gon-CHY}_{\rm sym} &=\frac{1}{\ell^2\,
(\ell+ k_{1})^2 (\ell+k_{1}+k_{2})^2\ldots  (\ell- k_{n})^2 }+{\rm Permutations\,\,}\{1,2,\ldots,n \}
\\
&=
 \frac{1}{\ell^2} \sum_{\a\in S_n}\frac{1}{
(\ell+ k_{\a_1})^2 (\ell+k_{\a_1}+k_{\a_2})^2\ldots  (\ell- k_{\a_n})^2 }
\nonumber\\
&= {\cal I}^{\rm n-gon-FEY}_{\rm sym},\nonumber
\end{align}
where $\a_i:=\a(i)$ and $S_n$ is the permutation group.

{\it  Remark.}\\
{\it Notice that we have been able to obtain the  $\Phi^3$ quadratic Feynman integrand at one-loop from the  CHY approach at tree-level and for massless particles.
In addition, clearly the  CHY-graphs in figures \ref{CHY_prescription} can be obtained from the one given in figure \ref{CHY-pgon-sym},  just by splitting the loop punctures in two on-shell particles. This is just a superficial fact, because the most interesting geometric consequences are going to be discussed in section  \ref{conclusion}.}

%%%%%%%%%%%%%%%%%%%%%%
\subsection*{General Diagram}
%%%%%%%%%%%%%%%%%%%%%%

\begin{figure}[!h]
  \centering
         \includegraphics[width=2in]{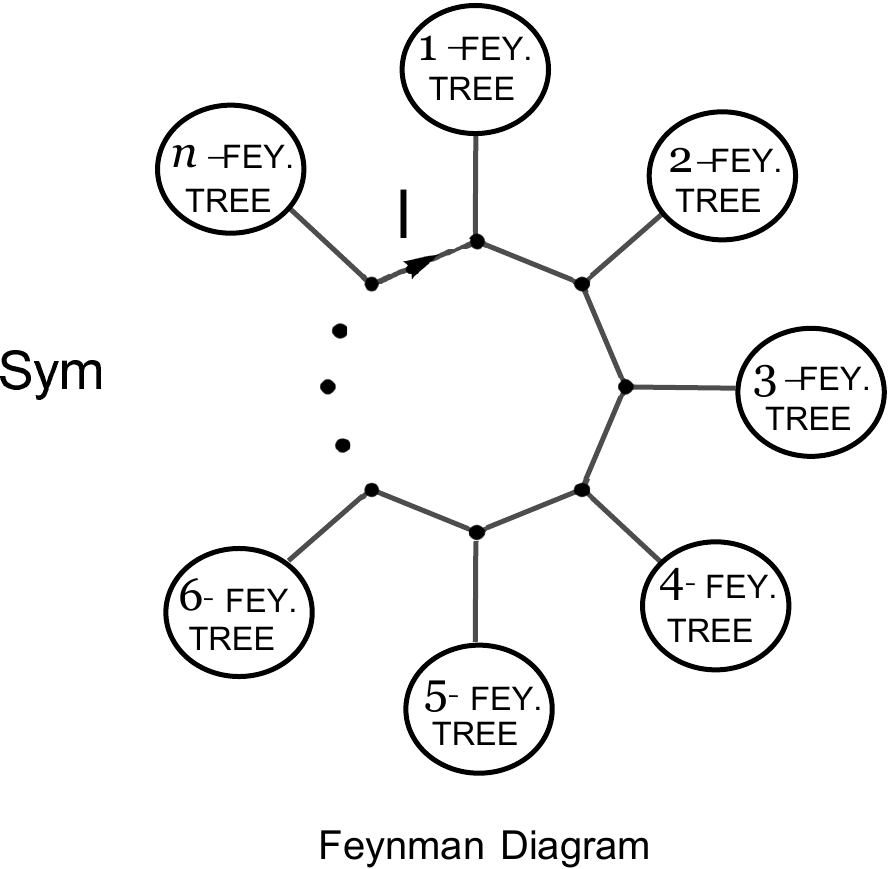}     \quad    
         \raisebox{2.1\height}{\includegraphics[scale=.5]{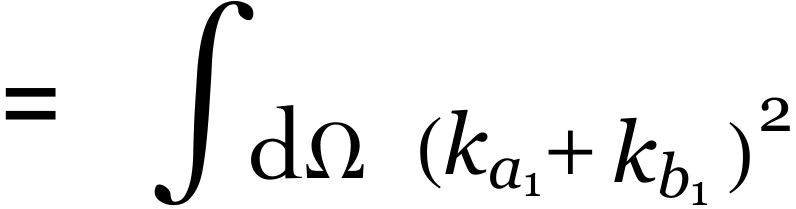}} \,\,\,\,                                     
         \includegraphics[width=1.8in]{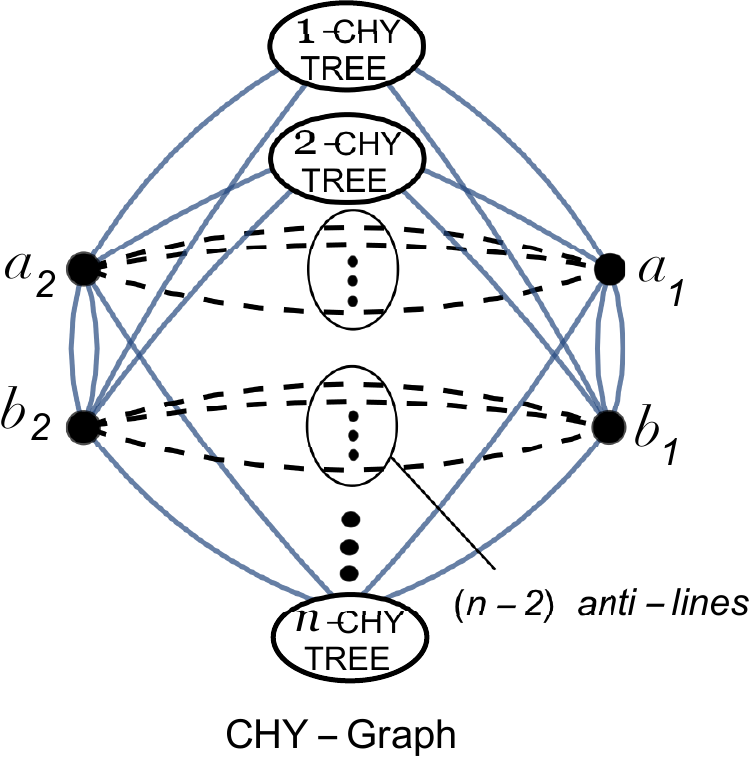}
  \caption{Schematic equivalence  between the $\Phi^3$ general Feynman diagrams (symmetrized $n$-gon)  and the CHY graphs.}\label{Equi-Fey-Chy}
\end{figure}

Using the techniques presented by the author {\it et al.} in \cite{Cardona:2016wcr,Gomez:2016cqb}, the generalization for any $\Phi^3$ Feynman diagram is very simple. Schematically, the equivalence among any  $\Phi^3$ Feynman diagram at one-loop and its corresponding CHY-graph is given in figure \ref{Equi-Fey-Chy}, where the symbol, $\mathsf{Sym}$, means symmetrization, namely a sum over all permutations of external legs.
%%%%%%%%%%%%%%%%%%%
%%%%%%%%%%%%%%%%%%%%%%%%%%%%%
\noindent

As it was argued above, the equivalence in figure \ref{Equi-Fey-Chy} is in fact an equality and the partial fraction identity is not necessary anymore. We will give a simple example in the next section.

%%%%%%%%%%%%%%%%%%%%%%%%
\section{Examples}\label{examples}
%%%%%%%%%%%%%%%%%%%%%%%%

In this section we consider three examples in order to check 
and illustrate the $\mathfrak{I}^{\rm n-gon-CHY}_{\rm sym}$ formula. We begin with the simplest one-loop case, the bubble. Latter, we will compute the triangle and finally an example of four-particle  will be given, where we are going to use the  generalization schematized in figure \ref{Equi-Fey-Chy}. 

%%%%%%%%%%%%%%%%%%%%%%%%
\subsection{The Bubble}\label{bubblesec}
%%%%%%%%%%%%%%%%%%%%%%%%

\begin{figure}[!h]
  \centering
    \includegraphics[width=1.3in]{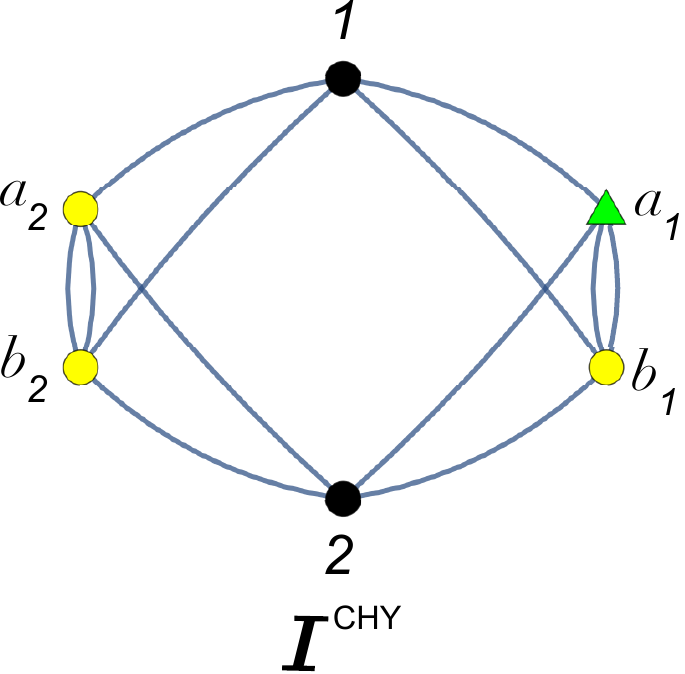} \\                                              
      \includegraphics[width=1.25in]{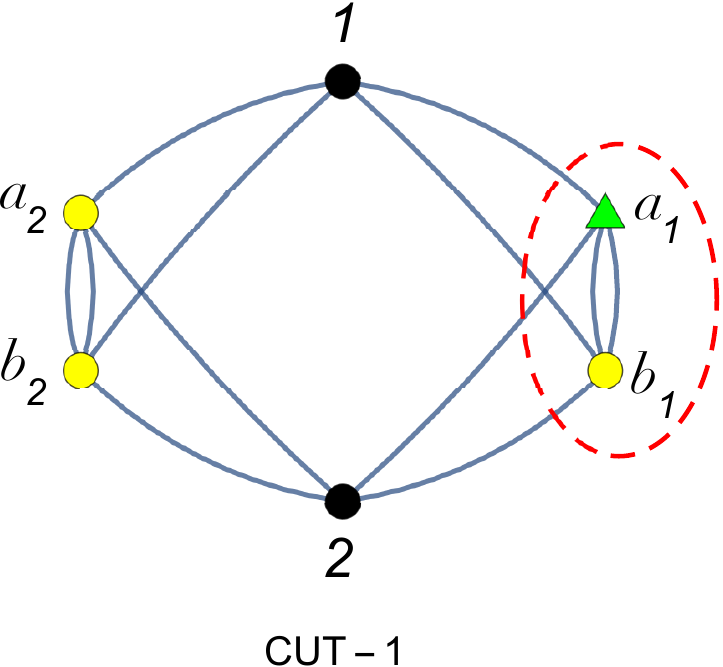}
        \includegraphics[width=1.25in]{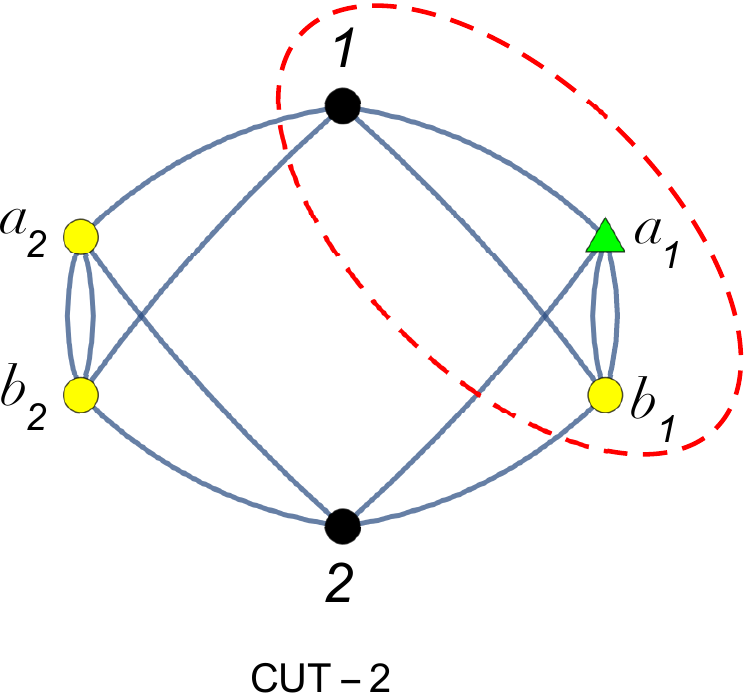}
          \includegraphics[width=1.25in]{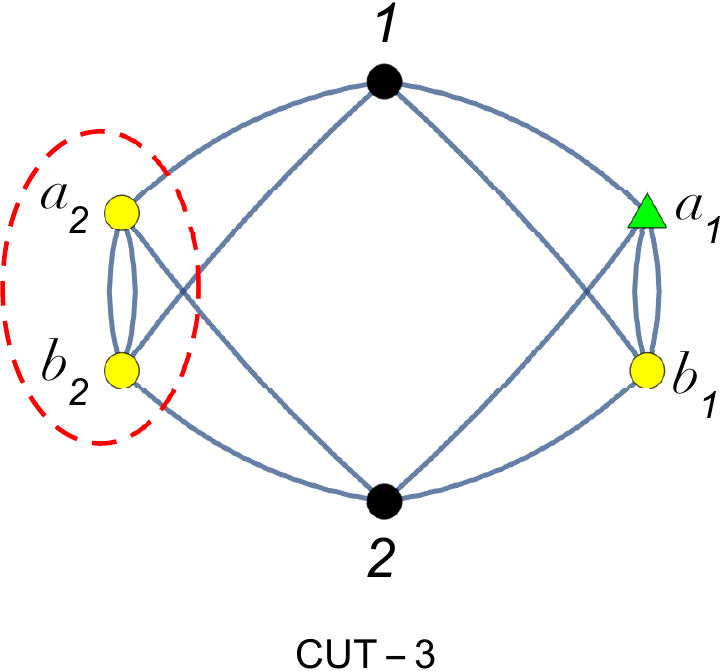}
            \includegraphics[width=1.25in]{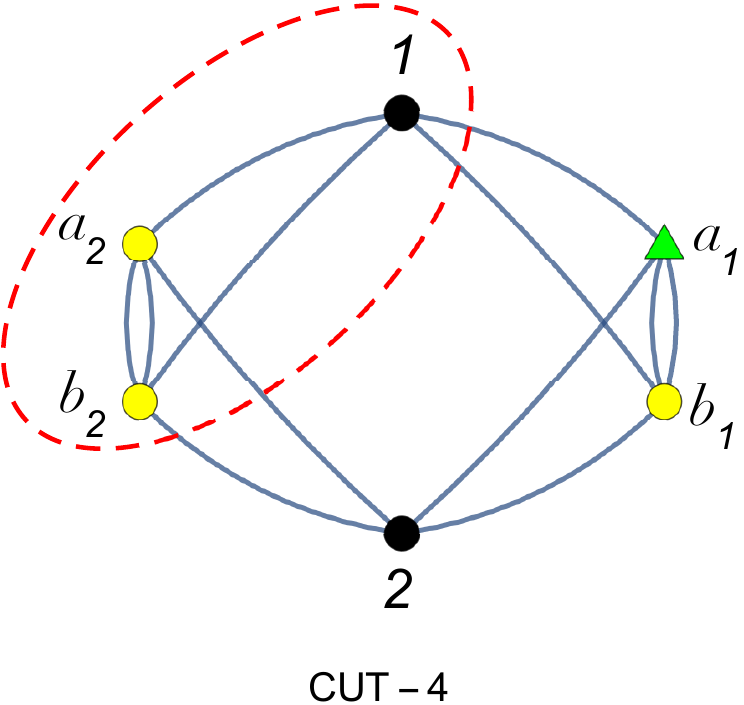}
  \caption{CHY-graph representation of ${\bf I}^{\rm CHY}$ written in \eqref{CHY-I-bubble}. All possible non-zero cuts for ${\bf I}^{\rm CHY}$.}\label{bubble-graph}
\end{figure}

Let us consider the $\mathbf{I}^{\rm CHY}$ integrand  given by expression
\begin{equation}\label{CHY-I-bubble}
\mathbf{I}^{\rm CHY} = 
\frac{1}{(a_1,b_1,b_2,a_2)^2}\,\,(\o^{a_1:a_2}_{1:2}\,\o^{a_1:a_2}_{2:1})\times (\o^{b_1:b_2}_{1:2}\,\o^{b_1:b_2}_{2:1}).
\end{equation}
Its  CHY-graph is represented on the top in figure \ref{bubble-graph}.
%%%%%%%%%%%%%%%%%%%%%%%%%%%%%%%%%%%
%%%%%%%%%%%%%%%%%%%%%%%%%%%%%
\noindent

To perform the integral, $\int d\mu\,\mathbf{I}^{\rm CHY}$,  we  apply the $\L-$algotihm developed in \cite{Gomez:2016bmv}. So, the all possible non-zero cuts have been  drawn on the second line in figure \ref{bubble-graph}. 

\begin{figure}[!h]
  \centering
    \includegraphics[width=1.5in]{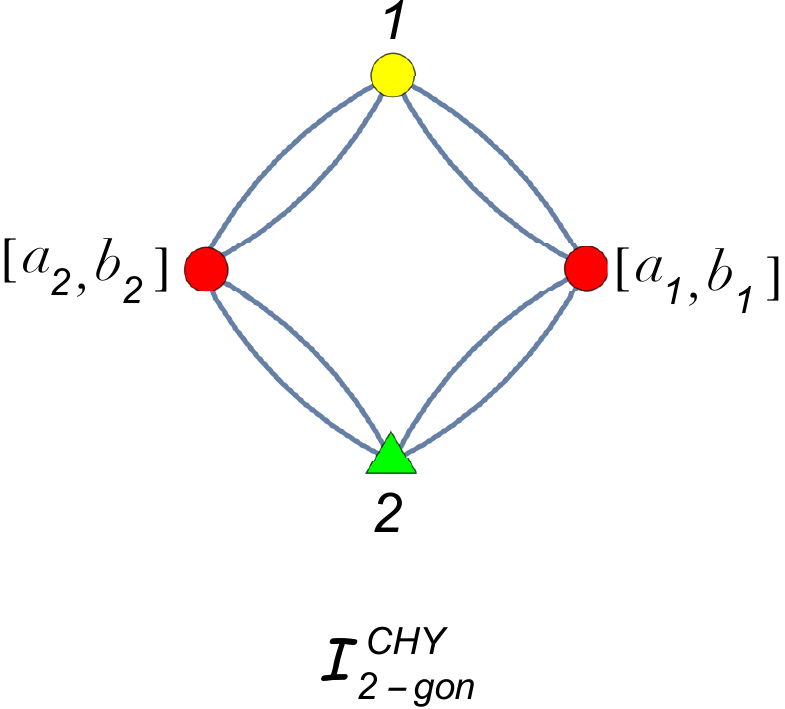}                                              
  \caption{Graph representation for ${\cal I}^{\rm 2-gon-CHY}_{\rm sym}(1,2 | [a_1,b_1],[a_2,b_2])$ (up to integration over the Moduli space, to wit $\int d\mu$).}\label{chy-2-gon}
\end{figure}

The computation of each cut is simple and the final answer for each one of them is
\begin{align*}
\mathbf{I}^{\rm CHY}_{\rm cut-1}=&\frac{1}{k_{a_1b_1}}\left[ \frac{1}{k_1\cdot (k_{a_1}+k_{b_1})}\times \frac{1}{k_2\cdot (k_1+k_{a_1}+k_{b_1})}  +\frac{{\cal I}^{\rm 2-gon-CHY}_{\rm sym}(1,2 | [a_1,b_1],[a_2,b_2])}{(k_1+k_2)\cdot(k_{a_1}+k_{b_1})+k_{12}} 
\right],\\
\mathbf{I}^{\rm CHY}_{\rm cut-2}=&\frac{1}{k_{1a_1b_1}}\times  \frac{1}{k_2\cdot (k_1+k_{a_1}+k_{b_1})}\times \frac{1}{k_1\cdot (k_2+k_{a_2}+k_{b_2})} , \\
\\
\mathbf{I}^{\rm CHY}_{\rm cut-3}=&\mathbf{I}^{\rm CHY}_{\rm cut-1}\Big|_{a_1\leftrightarrow a_2 \atop b_1\leftrightarrow b_2},\qquad
\mathbf{I}^{\rm CHY}_{\rm cut-4}=\mathbf{I}^{\rm CHY}_{\rm cut-2}\Big|_{a_1\leftrightarrow a_2 \atop b_1\leftrightarrow b_2} ,
\end{align*}
where  ${\cal I}^{\rm 2-gon-CHY}_{\rm sym}(1,2 | [a_1,b_1],[a_2,b_2])$ is the computation of the  CHY-graph drawn in figure \ref{chy-2-gon}.
%%%%%%%%%%%%%%%%%%%%%%%%%%%%%%%%%%%
%%%%%%%%%%%%%%%%%%%%%%%%%%%%%
\noindent
\\
Note that when $k_{a_2}+k_{b_2}=-(k_{a_1}+k_{b_1})$, namely on the support $\delta^{(D)} (k_{a_2}+k_{a_1})$ and $\delta^{(D)} (k_{b_2}+k_{b_1})$,  the graph in figure \ref{chy-2-gon} becomes one given in figure \ref{CHY-pgon-sym}. However,  on this support the denominator, $(k_1+k_2)\cdot(k_{a_1}+k_{b_1})+k_{12}$, vanishes, i.e. there is a singularity. In fact,  this is a spurious singularity\footnote{At two-loop this is a physical pole, which is related  with singularities of  higher codimension on a Riemann surface of genus two. Nevertheless, as it was shown in \cite{Gomez:2016cqb}, at two-loop these kind of contributions cancel out. It is going to be explained in section \ref{conclusion}.}, which can be removed using the momentum conservation condition, such as we are going to show. By summing, $\mathbf{I}^{\rm CHY}_{\rm cut-1}+\mathbf{I}^{\rm CHY}_{\rm cut-3}$,   one obtains  
\begin{align}\label{cut1_cut3}
&\mathbf{I}^{\rm CHY}_{\rm cut-1}+\mathbf{I}^{\rm CHY}_{\rm cut-3}=\\
&\frac{{\cal I}^{\rm 2-gon-CHY}_{\rm sym}(1,2 | [a_1,b_1],[a_2,b_2])}{[(k_1+k_2)\cdot(k_{a_1}+k_{b_1})+k_{12}]}
\frac{2}{(k_{a_1}+k_{b_1})^2} +
\frac{{\cal I}^{\rm 2-gon-CHY}_{\rm sym}(1,2 | [a_1,b_1],[a_2,b_2])}{[(k_1+k_2)\cdot(k_{a_2}+k_{b_2})+k_{12}]}
\frac{2}{(k_{a_2}+k_{b_2})^2} +\cdots\nonumber
\end{align}
where  $`` \cdots "$ means there are more terms. From the  momentum conservation condition, $k_1+k_2+k_{a_1}+k_{b_1}+k_{a_2}+k_{b_2}=0$, it is easy to check
\begin{align}\label{mcc_1}
&(k_1+k_2)\cdot(k_{a_1}+k_{b_1})+k_{12} = - (k_{a_1}+k_{b_1}+k_{a_2}+k_{b_2})\cdot (k_{a_1}+k_{b_1}) + \frac{1}{2}(k_{a_1}+k_{b_1}+k_{a_2}+k_{b_2})^2  \nonumber\\
&= \frac{1}{2} \left[ (k_{a_2}+k_{b_2})^2 - (k_{a_1}+k_{b_1})^2    \right]
\end{align}
and 
\begin{align}\label{mcc_2}
&(k_1+k_2)\cdot(k_{a_2}+k_{b_2})+k_{12} = - (k_{a_1}+k_{b_1}+k_{a_2}+k_{b_2})\cdot (k_{a_2}+k_{b_2}) + \frac{1}{2}(k_{a_1}+k_{b_1}+k_{a_2}+k_{b_2})^2  \nonumber\\
&= -\frac{1}{2} \left[ (k_{a_2}+k_{b_2})^2 - (k_{a_1}+k_{b_1})^2    \right].
\end{align}
Therefore, the sum in \eqref{cut1_cut3} becomes
\begin{align}
\mathbf{I}^{\rm CHY}_{\rm cut-1}+\mathbf{I}^{\rm CHY}_{\rm cut-3}&=\frac{2\,\,{\cal I}^{\rm 2-gon-CHY}_{\rm sym}(1,2 | [a_1,b_1],[a_2,b_2])}{(k_{a_2}+k_{b_2})^2 - (k_{a_1}+k_{b_1})^2}
\left [
\frac{2}{(k_{a_1}+k_{b_1})^2} -\frac{2}{(k_{a_2}+k_{b_2})^2} 
\right]+\cdots\nonumber\\
&=
\frac{4\,\,{\cal I}^{\rm 2-gon-CHY}_{\rm sym}(1,2 | [a_1,b_1],[a_2,b_2])}{(k_{a_1}+k_{b_1})^2\,\,(k_{a_2}+k_{b_2})^2}
+\cdots
\end{align}
Clearly, the spurious pole on the support, $ \delta^{(D)} ( k_{a_2}+k_{a_1}) \,\,\delta^{(D)} (k_{b_2}+k_{b_1} ) $,  has been cleanly removed\footnote{For higher number of points this mechanism works in the same way and these type of spurious singularities always are removed.}  and we can now compute the integral $\int d\Omega$. 

\begin{figure}[!h]
  \centering
         \includegraphics[width=1.4in]{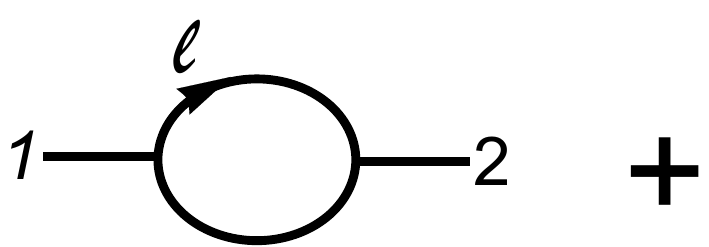}                                               
   \quad      \includegraphics[width=1.0in]{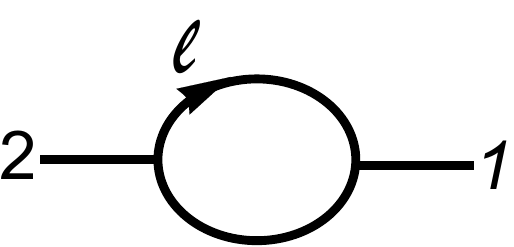}
  \caption{Feynman diagram for the symmetrized Bubble .}\label{bubble-fey}
\end{figure}

After removing the spurious pole and using the support, $ \delta^{(D)} ( k_{a_2}+k_{a_1}) \,\,\delta^{(D)} (k_{b_2}+k_{b_1} ) $, it is straightforward to check\footnote{Note that on the support, $ \delta^{(D)} ( k_{a_2}+k_{a_1}) \,\,\delta^{(D)} (k_{b_2}+k_{b_1} )$, the term 
\begin{equation}
{\cal I}^{\rm 2-gon-CHY}_{\rm sym}(1,2 | [a_1,b_1],[a_2,b_2])=\frac{1}{k_1\cdot(k_{a_1}+k_{b_1})}+\frac{1}{k_1\cdot(k_{a_2}+k_{b_2})}
\end{equation}
vanishes trivialy \cite{Cardona:2016bpi}. }
\begin{align}
&\mathbf{I}^{\rm CHY}_{\rm cut-1}+\mathbf{I}^{\rm CHY}_{\rm cut-2} +\mathbf{I}^{\rm CHY}_{\rm cut-3}+\mathbf{I}^{\rm CHY}_{\rm cut-4}=\frac{2}{k_{a_1b_1}k_{1a_1b_1} k_{2a_1b_1}}\\
&= \frac{2^4}{(k_{a_1}+k_{b_1})^2 (k_{1}+k_{a_1}+k_{b_1})^2(k_{2}+k_{a_1}+k_{b_1})^2} \nonumber\\
&= \frac{2^3}{(k_{a_1}+k_{b_1})^2} \left[  \frac{1}{(k_{a_1}+k_{b_1})^2(k_{a_1}+k_{b_1}+k_1)^2}
+\frac{1}{(k_{a_1}+k_{b_1})^2(k_{a_1}+k_{b_1}+k_2)^2}
\right].\nonumber
\end{align}
So, by computing the integral $\int d\Omega$, one obtains
\begin{align}
\mathfrak{I}^{\rm 2-gon-CHY}_{\rm sym} =&\int d\Omega\times \frac{(k_{a_1}+k_{b_1})^2}{2^3}\times 
\left(
\mathbf{I}^{\rm CHY}_{\rm cut-1}+\mathbf{I}^{\rm CHY}_{\rm cut-2} +\mathbf{I}^{\rm CHY}_{\rm cut-3}+\mathbf{I}^{\rm CHY}_{\rm cut-4}\right)\\
=&\frac{1}{\ell^2(\ell+k_1)^2}
+\frac{1}{\ell^2(\ell+k_2)^2}\nonumber ,
\end{align}
which is the Feynman integrand of the sum of diagrams given in figure \ref{bubble-fey}, as it was  expected.
%%%%%%%%%%%%%%%%%%%%%%%%%%%%%%%%%%%
%%%%%%%%%%%%%%%%%%%%%%%%%%%%%
\noindent

%%%%%%%%%%%%%%%%%%%%%%%%%%
%%%%%%%%%%%%%%%%%%%%%%%%
\subsection{The Triangle}\label{trianglesec}
%%%%%%%%%%%%%%%%%%%%%%%%
%%%%%%%%%%%%%%%%%%%%%%%%%%

As a second example we are going to  consider the next case, the triangle.  

\begin{figure}[!h]
  \centering
           \includegraphics[width=1.5in]{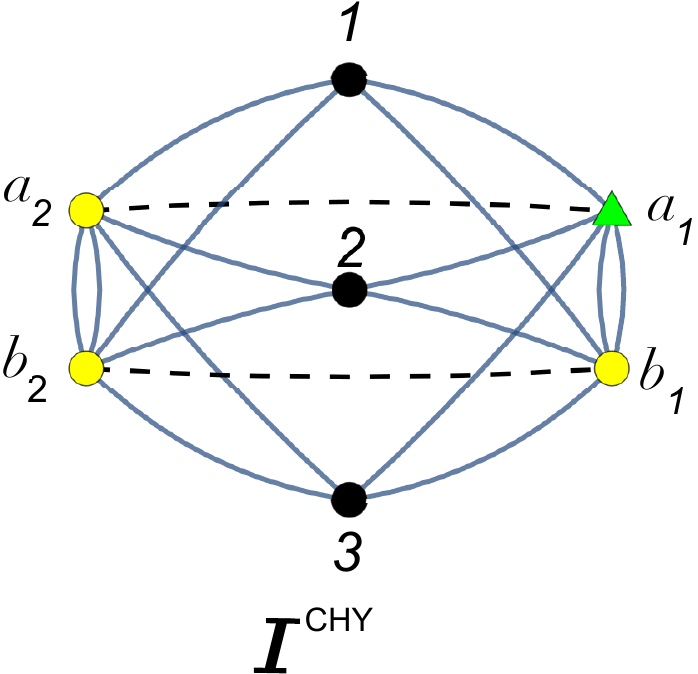}
  \caption{CHY-graph of the ${\bf I}^{\rm CHY}$ integrand given in \eqref{CHY-I-triangle}.}\label{CHY-triangle}
\end{figure}

Let $\mathbf{I}^{\rm CHY}$ be the CHY integrand  represented by the graph in figure \ref{CHY-triangle}
and given by the analytic expression 
\begin{equation}\label{CHY-I-triangle}
\mathbf{I}^{\rm CHY} = 
\frac{1}{(a_1,b_1,b_2,a_2)^2}\,\,(\o^{a_1:a_2}_{1:2}\,\o^{a_1:a_2}_{2:3}\,\o^{a_1:a_2}_{3:1})\times (\o^{b_1:b_2}_{1:3}\,\o^{b_1:b_2}_{3:2}\,\o^{b_1:b_2}_{2:1}).
\end{equation}
%%%%%%%%%%%%%%%%%%%%%%%%%%%%%%%%%%%
%%%%%%%%%%%%%%%%%%%%%%%%%%%%%
\noindent
\\

\begin{figure}[!h]
  \centering
         \includegraphics[width=1.4in]{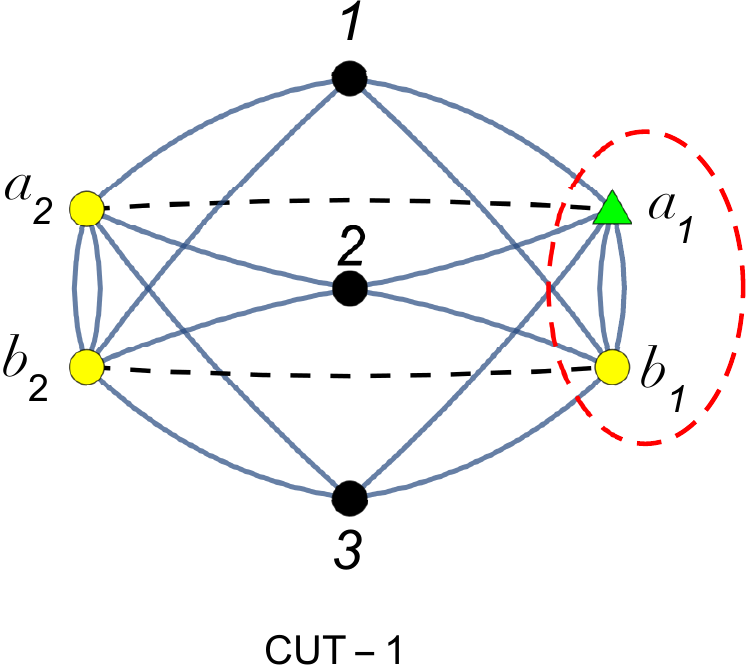}
         \includegraphics[width=1.4in]{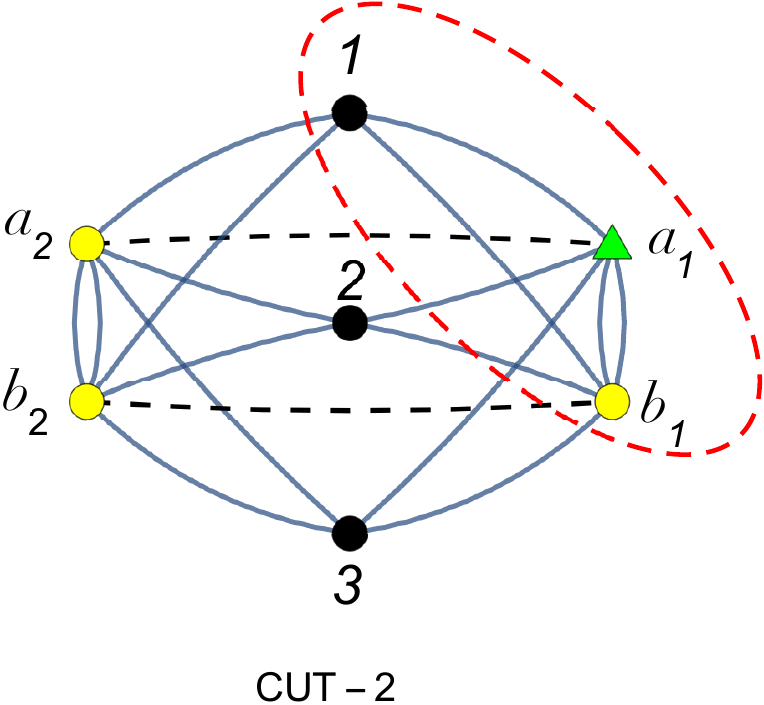}
         \includegraphics[width=1.4in]{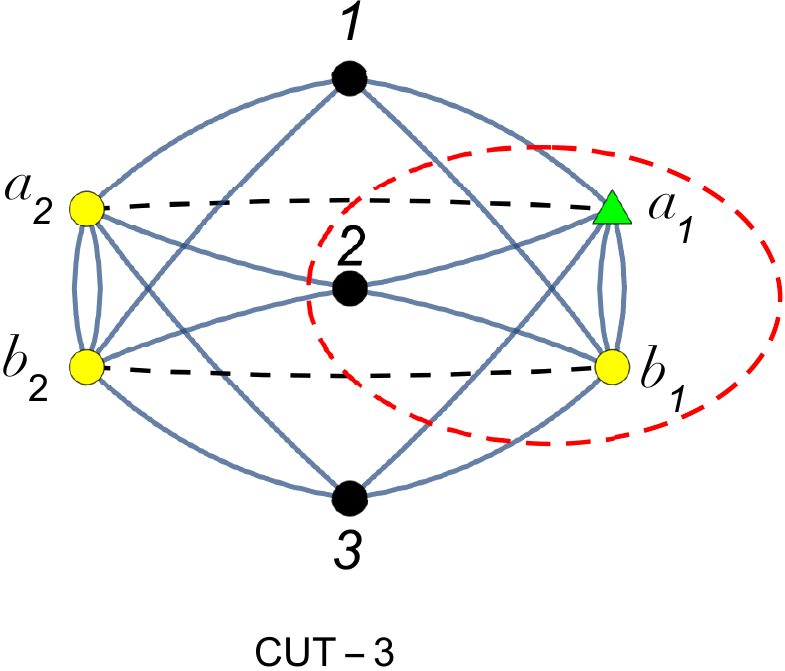}
            \includegraphics[width=1.4in]{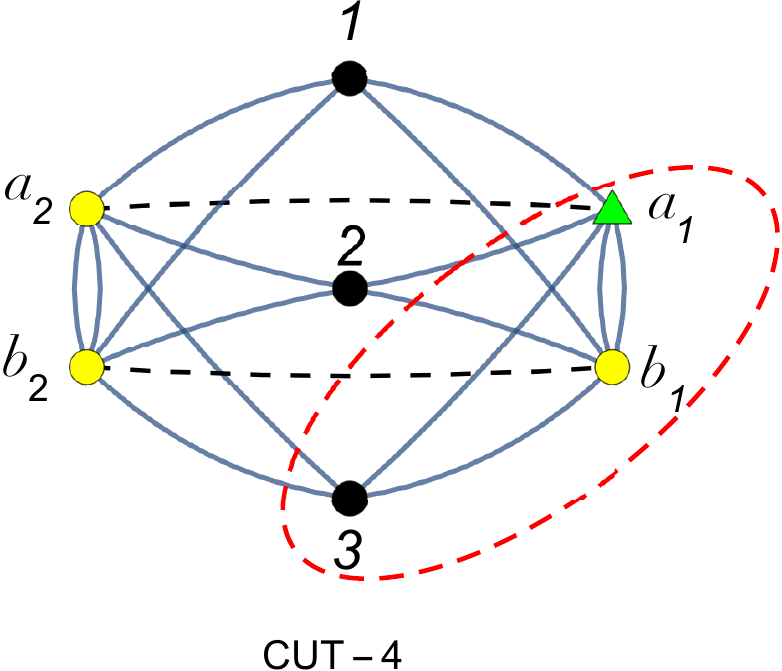}\\
     \includegraphics[width=1.4in]{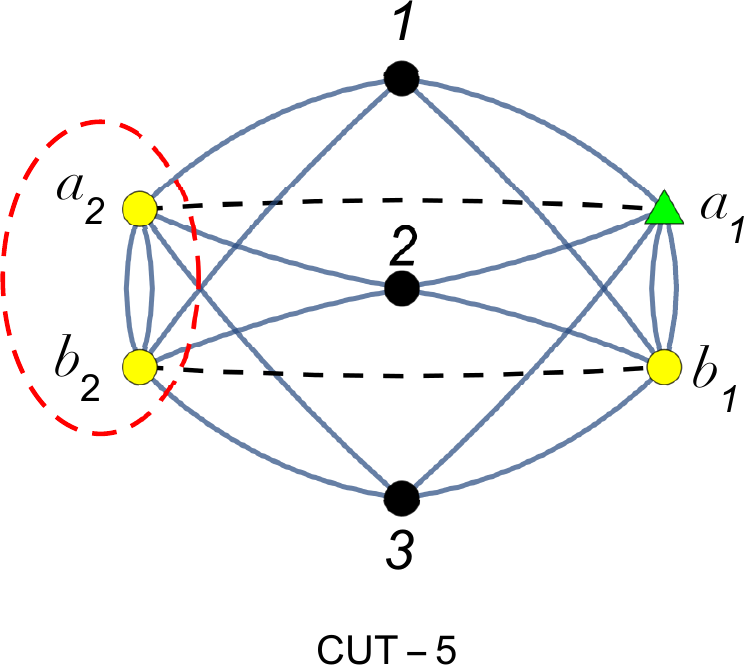}
       \includegraphics[width=1.4in]{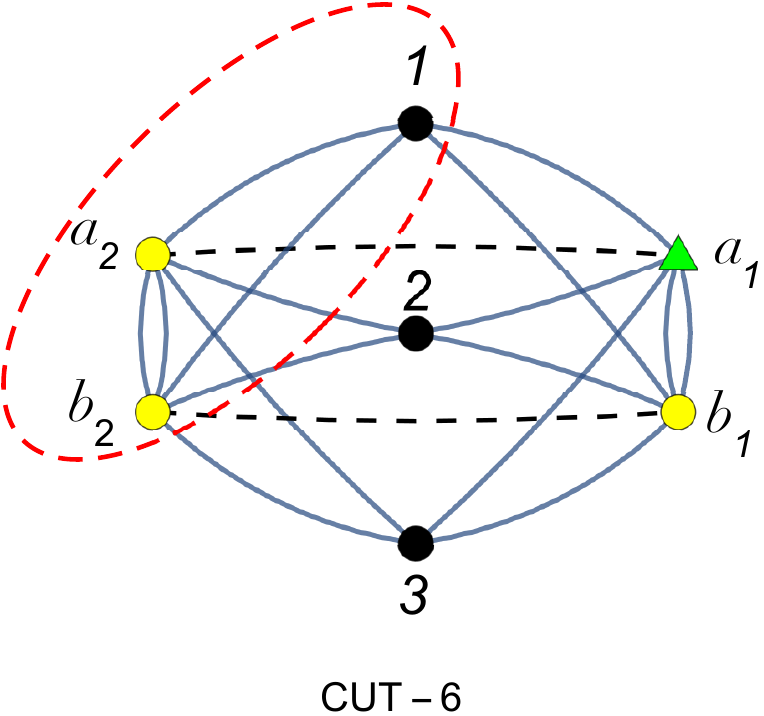}
         \includegraphics[width=1.4in]{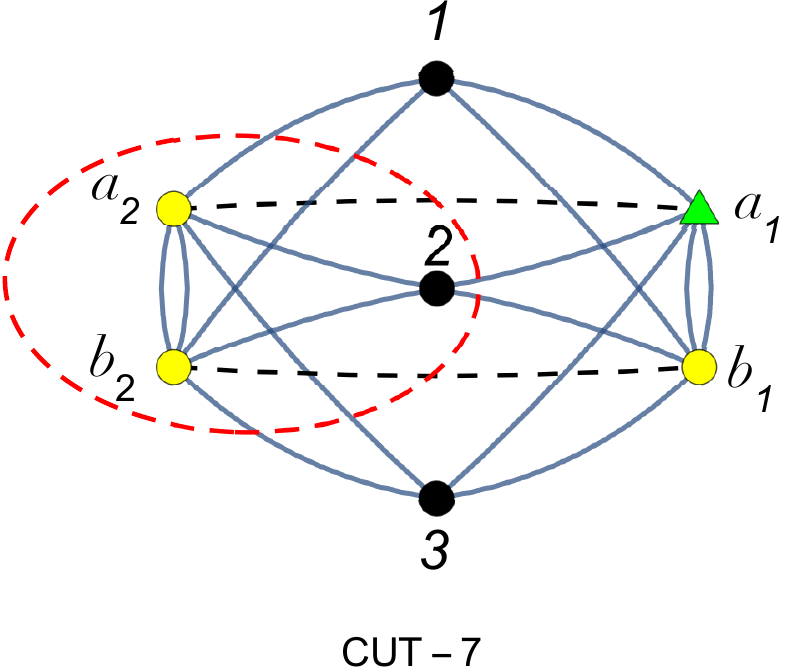}
            \includegraphics[width=1.4in]{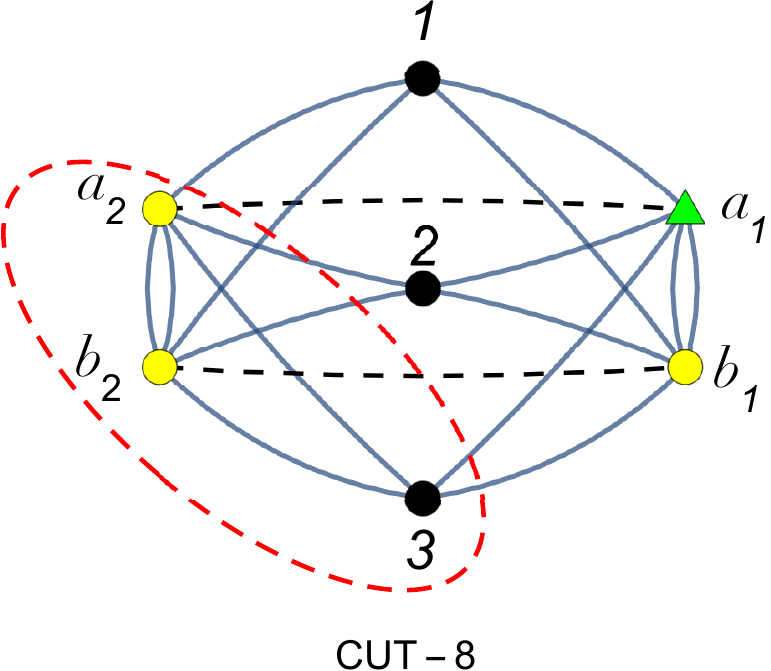}
  \caption{All possible non-zero cuts.}\label{CHY-triangle-cuts}
\end{figure}

In order to compute $\int d\mu \,\,\mathbf{I}^{\rm CHY}$, we  use the $\L-$algorithm. So, in figure \ref{CHY-triangle-cuts}  we have drawn the all possible non-zero cuts.
%%%%%%%%%%%%%%%%%%%%%%%%%%%%%
\noindent
\\

\begin{figure}[!h]
  \centering
         \includegraphics[width=1.9in]{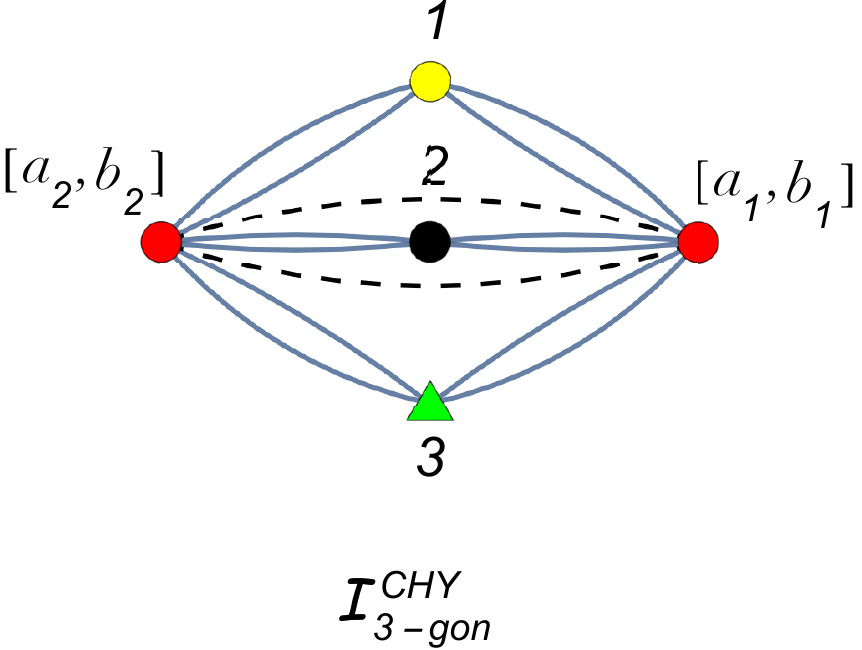}
  \caption{CHY-graph representation for ${\cal I}^{\rm 3-gon-CHY}_{\rm sym}(1,2,3 | [a_1,b_1],[a_2,b_2])$, up to integral over the Moduli space, i.e. $\int d\mu$.}\label{CHY-3gon}
\end{figure}

The computations are not hard and the final answer for each cut is
\begin{align*}
&\mathbf{I}^{\rm CHY}_{\rm cut-1}=\frac{1}{k_{a_1b_1}}\left\{ \frac{1}{k_1\cdot (k_{a_1}+k_{b_1})}\left[\frac{1}{k_2\cdot (k_1+k_{a_1}+k_{b_1})} \times \frac{1}{k_3\cdot(k_1+k_2+k_{a_1}+k_{b_1})}
\right.\right.
\\
&\left.
+
\frac{B( \{3\} , \{ 2\},\{1,a_1,b_1 \}, \{ a_2,b_2\}  )}{(k_2+k_3)(k_1+k_{a_1}+k_{b_1})+k_{23}} \right]
 +\frac{1}{(k_1+k_2)\cdot(k_{a_1}+k_{b_1})+k_{12}} \times \frac{B (\{2 \}, \{1 \}, \{a_1,b_1 \}, \{ 3,a_2,b_2 \}   )}{k_3\cdot(k_1+k_2+k_{a_1}+k_{b_1})}
\\
&
 +\frac{1}{(k_1+k_3)\cdot(k_{a_1}+k_{b_1})+k_{13}} \times \frac{B ( \{ 3\}, \{ 1\}, \{ a_1,b_1\}, \{2,a_2,b_2 \}  )}{k_2\cdot(k_1+k_3+k_{a_1}+k_{b_1})}
\left.
 +\frac{{\cal I}^{\rm 3-gon-CHY}_{\rm sym}(1,2,3 | [a_1,b_1],[a_2,b_2])}{(k_1+k_2+k_3)\cdot(k_{a_1}+k_{b_1})+k_{123}} 
\right\},
\end{align*}
\begin{align*}
&\mathbf{I}^{\rm CHY}_{\rm cut-2}=\frac{1}{k_{1a_1b_1}}\left\{
\frac{1}{k_1\cdot (k_2+k_3+k_{a_2}+k_{b_2})}\times
\left[ \frac{1}{k_2\cdot(k_1+k_{a_1}+k_{b_1})}\times  \frac{1}{k_3\cdot(k_1+k_2+k_{a_1}+k_{b_1})}
\right.\right.\\
&\left.\left.
+\frac{B (\{ 3\}, \{2 \}, \{ 1,a_1,b_1 \}, \{a_2,b_2 \}   ) }{(k_2+k_3)\cdot(k_1+k_{a_1}+k_{b_1})+k_{23}}
\right]
\right\},
\end{align*}
\begin{align*}
&\mathbf{I}^{\rm CHY}_{\rm cut-3}=\mathbf{I}^{\rm CHY}_{\rm cut-2}\Big|_{k_1\leftrightarrow k_2},\qquad
\mathbf{I}^{\rm CHY}_{\rm cut-4}=\mathbf{I}^{\rm CHY}_{\rm cut-2}\Big|_{k_1\leftrightarrow k_3},\qquad
\mathbf{I}^{\rm CHY}_{\rm cut-5}=\mathbf{I}^{\rm CHY}_{\rm cut-1}\Big|_{a_1\leftrightarrow a_2 \atop b_1\leftrightarrow b_2},\\
&\mathbf{I}^{\rm CHY}_{\rm cut-6}=\mathbf{I}^{\rm CHY}_{\rm cut-2}\Big|_{a_1\leftrightarrow a_2 \atop b_1\leftrightarrow b_2},\qquad
\mathbf{I}^{\rm CHY}_{\rm cut-7}=\mathbf{I}^{\rm CHY}_{\rm cut-3}\Big|_{a_1\leftrightarrow a_2 \atop b_1\leftrightarrow b_2},\qquad
\mathbf{I}^{\rm CHY}_{\rm cut-8}=\mathbf{I}^{\rm CHY}_{\rm cut-4}\Big|_{a_1\leftrightarrow a_2 \atop b_1\leftrightarrow b_2},
\end{align*}
where  ${\cal I}^{\rm 3-gon-CHY}_{\rm sym}(1,2,3 | [a_1,b_1],[a_2,b_2])$ is the computation of the  CHY-graph given in figure \ref{CHY-3gon}.
%%%%%%%%%%%%%%%%%%%%%%%%%%%%%
\noindent
Clearly, such as it happened in the previous example, there is a spurious pole when the support, $ \delta^{(D)} ( k_{a_2}+k_{a_1}) \,\,\delta^{(D)} (k_{b_2}+k_{b_1} )$, is considered, i.e. the cuts $\mathbf{I}^{\rm CHY}_{\rm cut-1}$ and  $\mathbf{I}^{\rm CHY}_{\rm cut-5}$ become infinite when $k_{a_2}=-k_{a_1}$ and $k_{b_2}=-k_{b_1}$. In order to remove it,  we apply the same trick as in section \ref{bubblesec}. So, by considering the sum, $\mathbf{I}^{\rm CHY}_{\rm cut-1} + \mathbf{I}^{\rm CHY}_{\rm cut-5}$,  and using the momentum conservation condition, such as in \eqref{mcc_1} and \eqref{mcc_2}, it is straightforward to check
\begin{align*}
&\mathbf{I}^{\rm CHY}_{\rm cut-1} + \mathbf{I}^{\rm CHY}_{\rm cut-5} = 
 \frac{ 4\,\,{\cal I}^{\rm 3-gon-CHY}_{\rm sym}(1,2,3 | [a_1,b_1],[a_2,b_2])}{(k_{a_1}+k_{b_1})^2\,\,(k_{a_2}+k_{b_2})^2}  + \ldots ,
\end{align*} 
and therefore the spurious singularity on  $ \delta^{(D)} ( k_{a_2}+k_{a_1}) \,\,\delta^{(D)} (k_{b_2}+k_{b_1} ) $ is removed.

\begin{figure}[!h]
  \centering
         \includegraphics[width=1.6in]{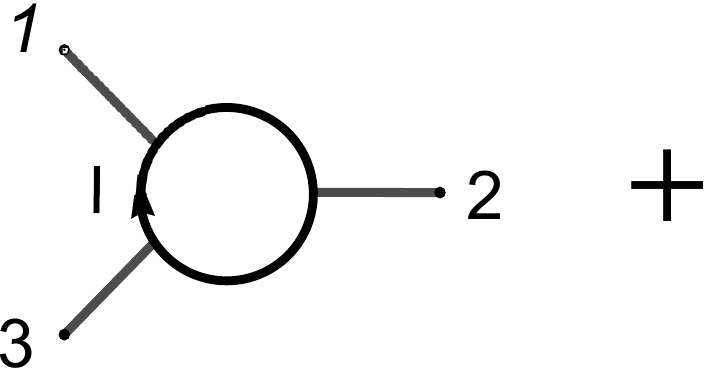}\qquad
             \raisebox{2.3\height}{\includegraphics[scale=0.55]{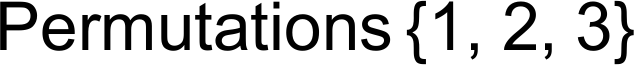}}
  \caption{Feynman diagram for the symmetrized Triangle.}\label{fey-triangle}
\end{figure}

Finally, after removing the spurious pole and on the support, $ \delta^{(D)} ( k_{a_2}+k_{a_1}) \,\,\delta^{(D)} (k_{b_2}+k_{b_1} ) $,  one can verify\footnote{Note that when, $ k_{a_2}=-k_{a_1}$ and $k_{b_2}=-k_{b_1}$,  the term 
\begin{align*}
{\cal I}^{\rm 3-gon-CHY}_{\rm sym}(1,2,3 | [a_1,b_1],[a_2,b_2])&=
\frac{B ( \{ 2\}, \{ 3 \} , \{ 1,a_1,b_1 \} ,\{ a_2,b_2 \})   }{(k_2+k_3)\cdot (k_{a_2}+k_{b_2})+k_{23}}
 + \frac{B ( \{ 2 \}, \{ 3 \} , \{ a_1,b_1 \} ,\{1, a_2,b_2 \})   }{(k_2+k_3)\cdot (k_{a_1}+k_{b_1})+k_{23}}  \\
&  +
\frac{B ( \{ 2 \}, \{ 1 \} , \{ 3,a_1,b_1 \} ,\{ a_2,b_2 \})   }{k_3\cdot (k_{a_1}+k_{b_1})}
 +  \frac{B ( \{ 2 \}, \{ 1 \} , \{ 3,a_2,b_2 \} ,\{ a_1,b_1 \})   }{k_3\cdot (k_{a_2}+k_{b_2})}
\end{align*}
vanishes trivialy \cite{Cardona:2016bpi}.}
\begin{align}
\sum_{i=1}^8 {\bf I}^{\rm CHY}_{{\rm cut - }i}
&=\frac{2^4}{s_{a_1b_1}} \left[ \frac{1}{s_{a_1b_1}\,(k_{a_1}+k_{b_1}+k_{1})^2\, (k_{a_1}+k_{b_1}+k_{1}+k_{2})^2}+{\rm Permutations}\,\,\{1,2,3 \} \right]\nonumber
\\
&=\frac{2^4}{s_{a_1b_1}} \sum_{\a\in S_3}\frac{1}{s_{a_1b_1}\,(k_{a_1}+k_{b_1}+k_{\a_1})^2\, (k_{a_1}+k_{b_1}+k_{\a_1}+k_{\a_2})^2}.
\end{align}
Therefore, by integrating $\int\,d\Omega$, it is simple to see
\begin{align}
\mathfrak{I}^{\rm 3-gon-CHY}_{\rm sym} &=\int d\Omega\times \frac{(k_{a_1}+k_{b_1})^2}{2^4}\times 
\left(\sum_{i=1}^8 {\bf I}^{\rm CHY}_{{\rm cut - }i}\right)\nonumber\\
&=\frac{1}{\ell^2 (\ell+k_{1})^2\, (\ell+k_{1}+k_{2})^2}+{\rm Permutations\,  \{ 1,2,3   \}}
\nonumber\\
&=\sum_{\a\in S_3}\frac{1}{\ell^2\,(\ell+k_{\a_1})^2\, (\ell+k_{\a_1}+k_{\a_2})^2}
,
\end{align}
which is the Feynman integrand of the six diagrams given in figure \ref{fey-triangle}, as it was  expected.
%%%%%%%%%%%%%%%%%%%%%%%%%%%%%
\noindent
\\

%%%%%%%%%%%%%%%%%%%%%%%%%%%%
%%%%%%%%%%%%%%%%%%%%%%%%
\subsection{Four-particle}\label{trianglesec}
%%%%%%%%%%%%%%%%%%%%%%%%
%%%%%%%%%%%%%%%%%%%%%%%%%%%%

In this section we look upon a more complicated example, a four-particle computation. In order to present an example with the same structure as in figure \ref{Equi-Fey-Chy}, we consider a  triangle with four-particle.

\begin{figure}[!h]
  \centering
        \includegraphics[scale=0.5]{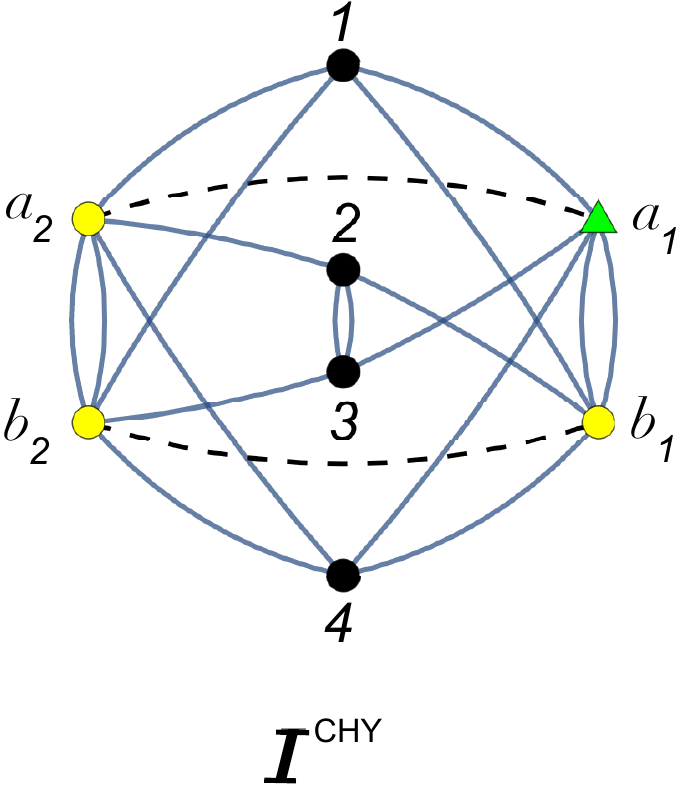}     
  \caption{CHY-graph  of the integrand in  \eqref{CHY-I-triangleF}}\label{4p_ex}
\end{figure}

Let ${\bf I}^{\rm CHY}$ be the CHY integrand given by 
\begin{equation}\label{CHY-I-triangleF}
\mathbf{I}^{\rm CHY} = 
\frac{1}{(a_1,b_1,b_2,a_2)^2}\,\,(\o^{a_1:a_2}_{1:2}\,\frac{1}{\s_{23}}\,\o^{a_1:a_2}_{3:4}\,\o^{a_1:a_2}_{4:1})\times (\o^{b_1:b_2}_{1:4}\,\o^{b_1:b_2}_{4:3}\,  \frac{1}{\s_{32}}\,  \o^{b_1:b_2}_{2:1}),
\end{equation}
and in figure \ref{4p_ex} we have represented its CHY-graph.
%%%%%%%%%%%%%%%%%%%%%%%%%%%%%
\noindent
\\

\begin{figure}[!h]
  \centering
         \includegraphics[width=1.3in]{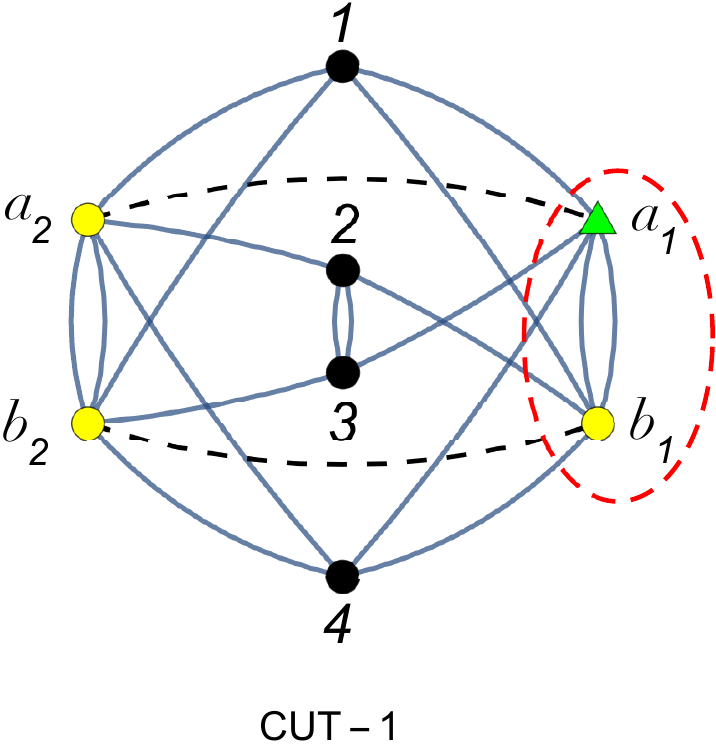}
         \includegraphics[width=1.3in]{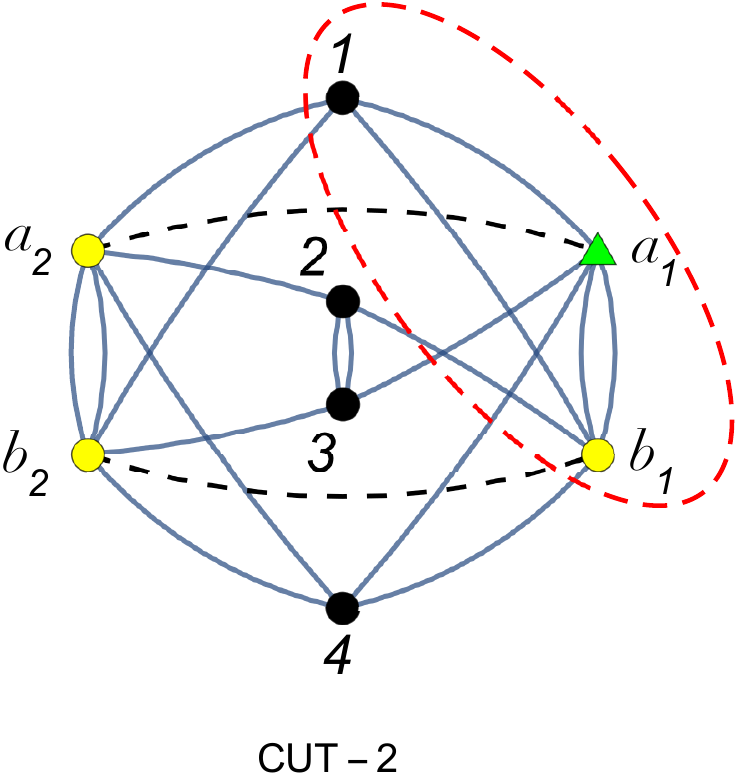}
         \includegraphics[width=1.3in]{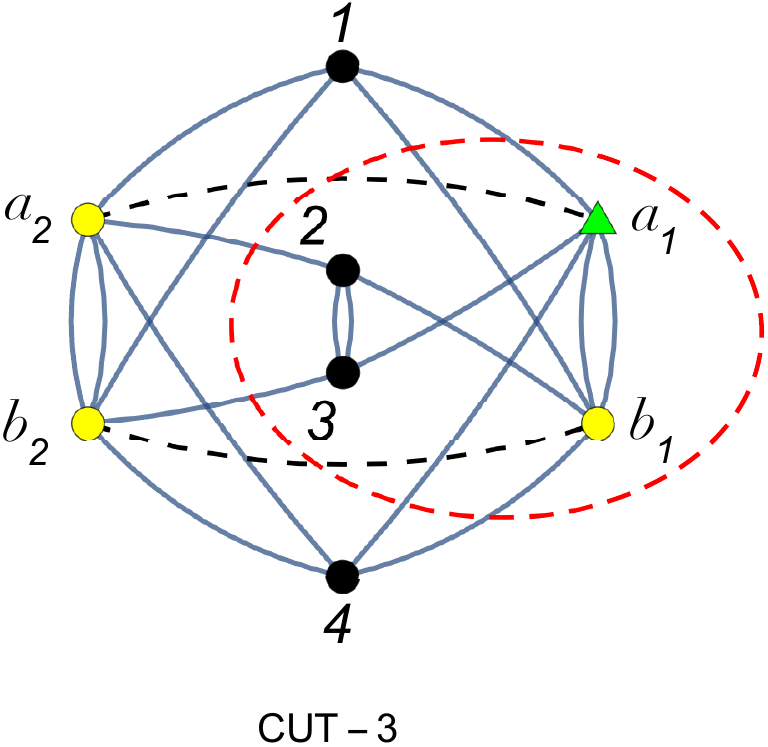}
            \includegraphics[width=1.3in]{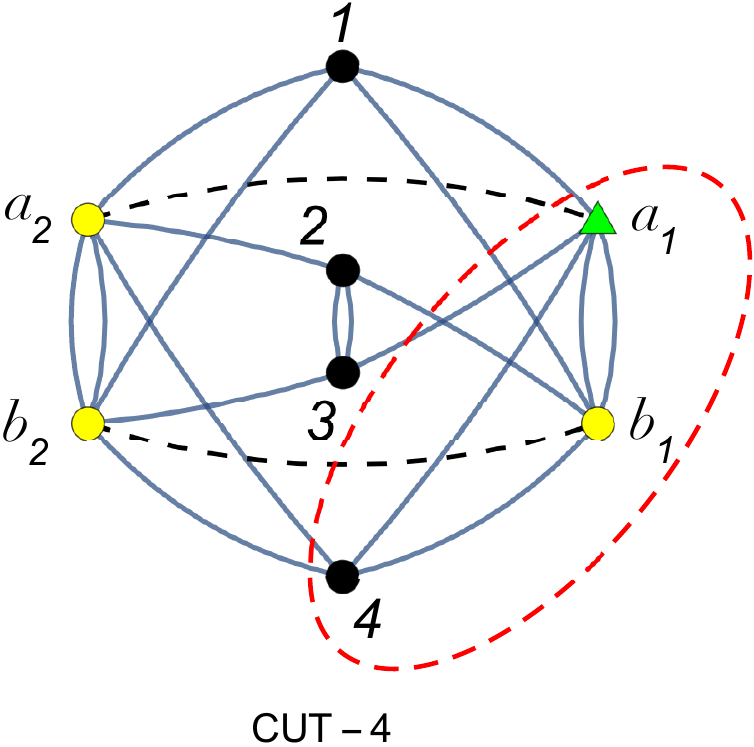}\\
     \includegraphics[width=1.3in]{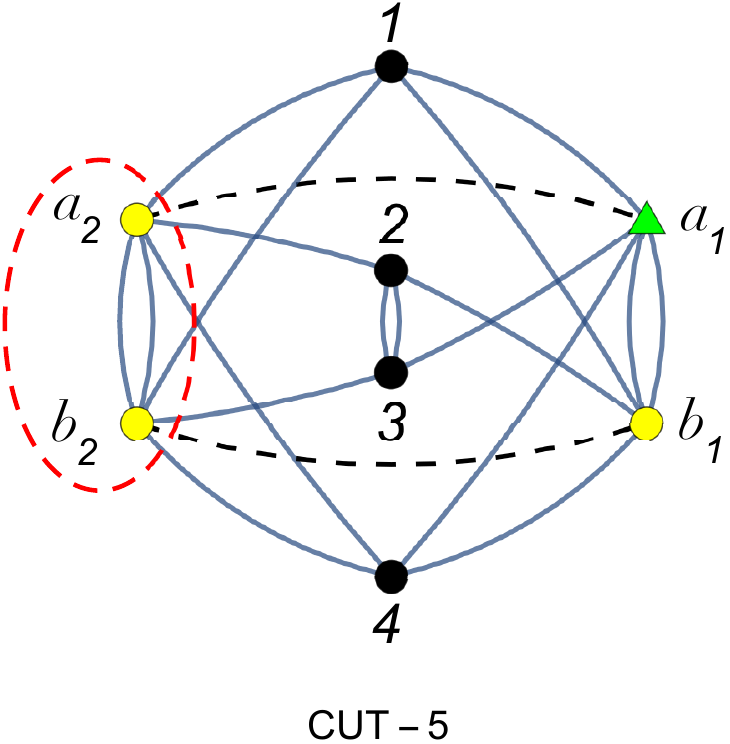}
       \includegraphics[width=1.3in]{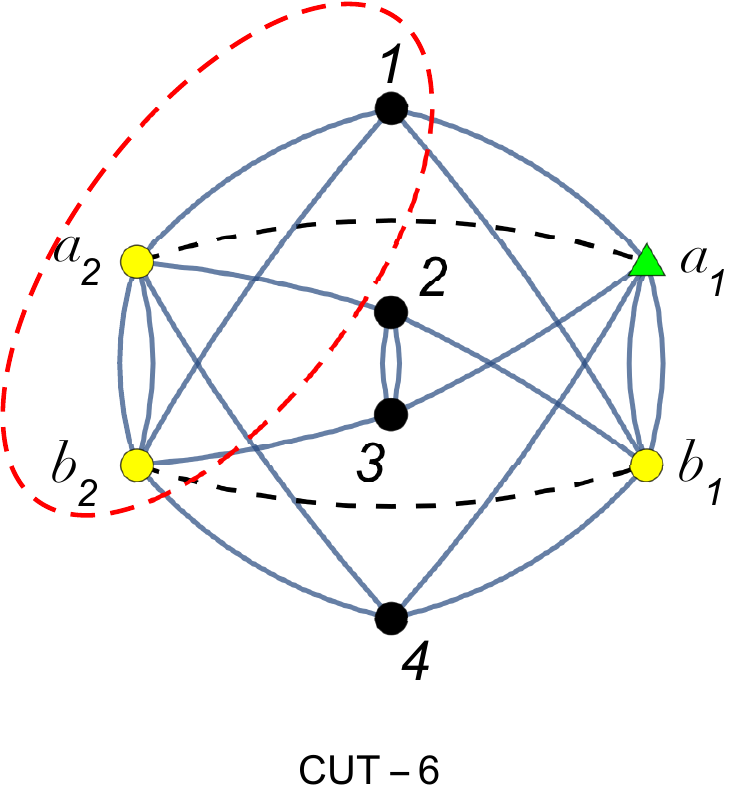}
         \includegraphics[width=1.3in]{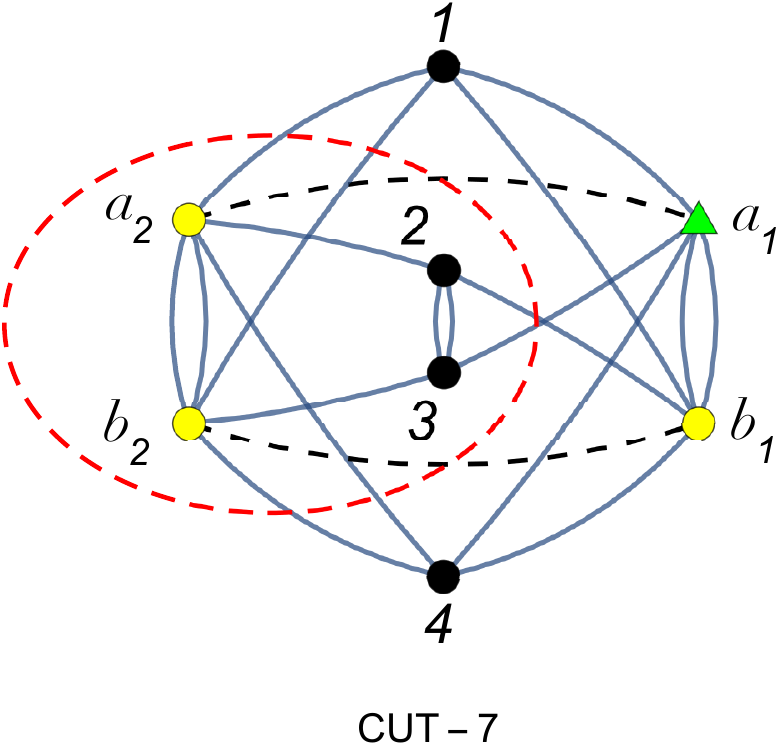}
            \includegraphics[width=1.3in]{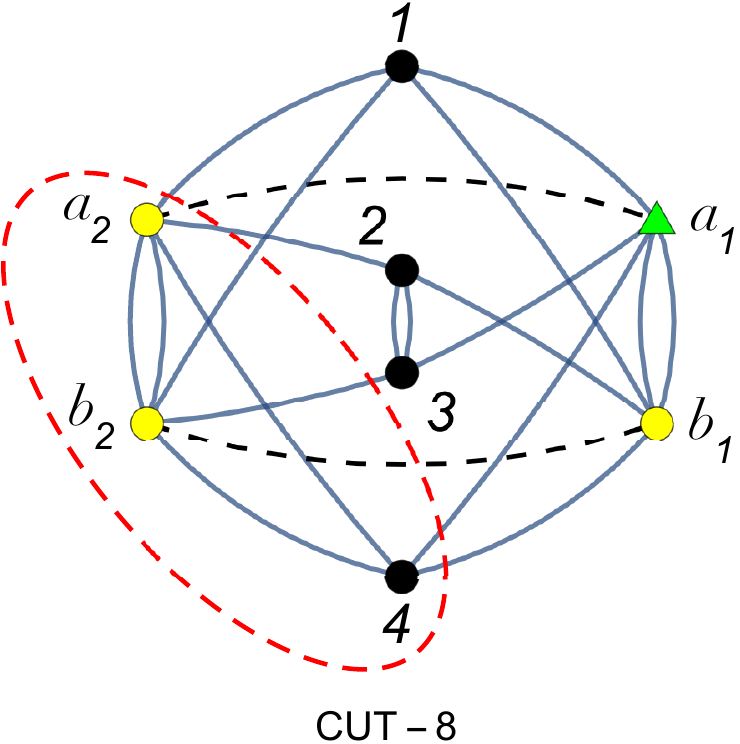}
  \caption{All possible non-zero cuts.}\label{CHY-4p-cuts}
\end{figure}

The computation of the integral, $\int d\mu\, {\bf I}^{\rm CHY}$, is performed by applying the $\L-$algorithm. Following this algorithm, we draw the all possible non-zero cuts in figure \ref{CHY-4p-cuts}.
%%%%%%%%%%%%%%%%%%%%%%%%%%%%%
\noindent
\\

\begin{figure}[!h]
  \centering
         \includegraphics[width=1.6in]{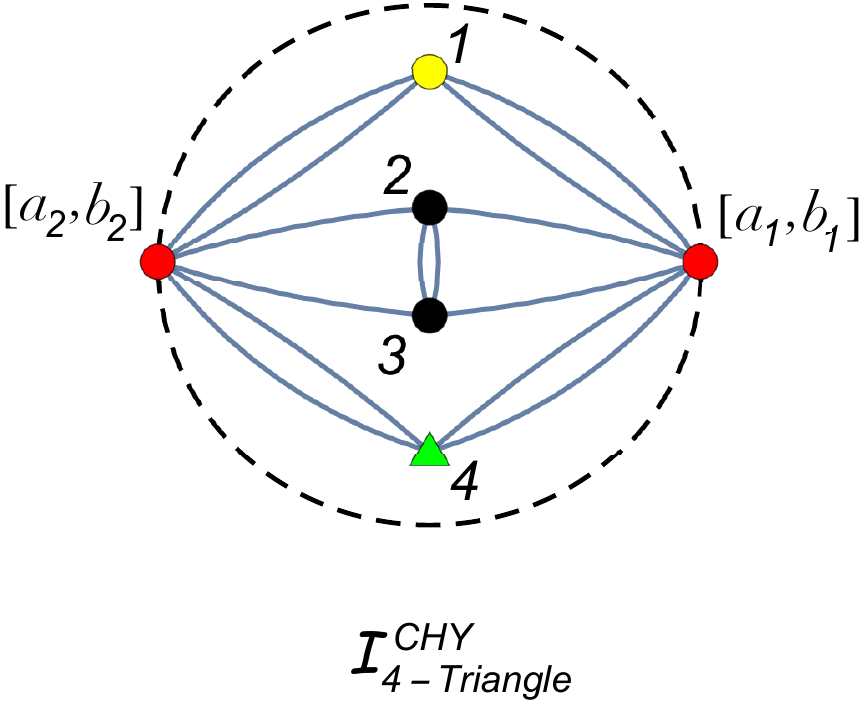}
  \caption{CHY-graph for ${\cal I}^{\rm CHY}_{\rm 4-Triangle}(1|2,3|4 | [a_1,b_1],[a_2,b_2])$, up to integral over the Moduli space, i.e. $\int d\mu$.}\label{CHY-4p-sing}
\end{figure}

The answer of each cut is obtained after a long but not hard computation and 
the final results are given  by the expressions
\begin{align*}
\mathbf{I}^{\rm CHY}_{\rm cut-1}= & \frac{1}{k_{23}\,\,k_{a_1b_1}}\left\{ \frac{1}{k_1\cdot (k_{a_1}+k_{b_1})}\left[\frac{1}{(k_2+k_3)\cdot (k_1+k_{a_1}+k_{b_1})+k_{23}} \times \frac{1}{k_4\cdot(k_1+k_2+k_3+k_{a_1}+k_{b_1})}
\right.\right.
\\
&\left.
+
\frac{B(\{4\}, \{2,3 \}, \{1,a_1,b_1 \},\{a_2,b_2 \} )}{(k_2+k_3+k_4)(k_1+k_{a_1}+k_{b_1})+k_{234}}\right]
\nonumber \\
&
 +\frac{1}{(k_1+k_2+k_3)\cdot(k_{a_1}+k_{b_1})+k_{123}} \times \frac{B( \{ 2,3 \},\{ 1 \},\{a_1,b_1 \},\{ 4,a_2,b_2 \}  )}{k_4\cdot(k_1+k_2+k_3+k_{a_1}+k_{b_1})}   \\
&
 +\frac{1}{(k_1+k_4)\cdot(k_{a_1}+k_{b_1})+k_{14}} \times \frac{B (\{ 4 \},\{1\}, \{ a_1,b_1 \},\{ 2,3,a_2,b_2 \})}{(k_2+k_3)\cdot(k_1+k_4+k_{a_1}+k_{b_1})+k_{23}}\\
& 
\left.
 +\frac{k_{23}\,\,{\cal I}^{\rm CHY}_{\rm 4-Triangle}(1|2,3| 4 | [a_1,b_1],[a_2,b_2])}{(k_1+k_2+k_3+k_4)\cdot(k_{a_1}+k_{b_1})+k_{1234}} 
\right\},
\end{align*}
\begin{align*}
\mathbf{I}^{\rm CHY}_{\rm cut-2}=&
\frac{1}{k_{23}\, k_{1a_1b_1}}\times \frac{1}{k_1\cdot (k_2+k_3+k_4+k_{a_2}+k_{b_2})}\times \\
&
\left[ \frac{1}{(k_2+k_3)\cdot(k_1+k_{a_1}+k_{b_1})+k_{23}}\times  \frac{1}{k_4\cdot(k_1+k_2+k_3+k_{a_1}+k_{b_1})} \right. \\
&\left.
+\frac{B( \{ 4 \},\{ 2,3 \},\{1,a_1,b_1 \}, \{  a_2,b_2 \} )}{(k_2+k_3+k_4)\cdot(k_1+k_{a_1}+k_{b_1})+k_{234}}
\right],
\end{align*}
\begin{align*}
\mathbf{I}^{\rm CHY}_{\rm cut-3}=&
\frac{1}{k_{23}\, k_{23a_1b_1}}\times \frac{1}{(k_2+k_3)\cdot (k_1+k_4+k_{a_2}+k_{b_2})+k_{23}}\times \\
&
\left[ \frac{1}{k_1\cdot(k_2+k_3+k_{a_1}+k_{b_1})}\times  \frac{1}{k_4\cdot(k_1+k_2+k_3+k_{a_1}+k_{b_1})} \right. \\
&\left.
+\frac{B( \{ 4 \},\{ 1\},\{2,3,a_1,b_1 \}, \{  a_2,b_2 \} )}{(k_1+k_4)\cdot(k_2+k_3+k_{a_1}+k_{b_1})+k_{14}}
\right],
\end{align*}
\begin{align*}
&
\mathbf{I}^{\rm CHY}_{\rm cut-4}=\mathbf{I}^{\rm CHY}_{\rm cut-2}\Big|_{k_1\leftrightarrow k_4},\qquad
\mathbf{I}^{\rm CHY}_{\rm cut-5}=\mathbf{I}^{\rm CHY}_{\rm cut-1}\Big|_{a_1\leftrightarrow b_2 \atop b_1\leftrightarrow a_2},\qquad
\mathbf{I}^{\rm CHY}_{\rm cut-6}=\mathbf{I}^{\rm CHY}_{\rm cut-2}\Big|_{a_1\leftrightarrow b_2 \atop b_1\leftrightarrow a_2}
\\
&
\mathbf{I}^{\rm CHY}_{\rm cut-7}=\mathbf{I}^{\rm CHY}_{\rm cut-3}\Big|_{a_1\leftrightarrow b_2 \atop b_1\leftrightarrow a_2},\qquad
\mathbf{I}^{\rm CHY}_{\rm cut-8}=\mathbf{I}^{\rm CHY}_{\rm cut-4}\Big|_{a_1\leftrightarrow b_2 \atop b_1\leftrightarrow a_2},
\end{align*}
where ${\cal I}^{\rm CHY}_{\rm 4-Triangle}(1|2,3| 4 | [a_1,b_1],[a_2,b_2]) $ is the computation of the CHY graph in figure \ref{CHY-4p-sing}.
%%%%%%%%%%%%%%%%%%%%%%%%%%%%%
\noindent
\\

As it happened in the above examples, the cuts, $\mathbf{I}^{\rm CHY}_{\rm cut-1}$ and $\mathbf{I}^{\rm CHY}_{\rm cut-5}$, have spurious pole on $\delta^{(D)} ( k_{a_2}+k_{a_1}) \,\,\delta^{(D)} (k_{b_2}+k_{b_1} )$.  In order to remove it, we consider the sum, $\mathbf{I}^{\rm CHY}_{\rm cut-1} + \mathbf{I}^{\rm CHY}_{\rm cut-5}$. Using the momentum conservation condition, such as in \eqref{mcc_1} and \eqref{mcc_2}, it is straightforward to see
\begin{align*}
&\mathbf{I}^{\rm CHY}_{\rm cut-1} + \mathbf{I}^{\rm CHY}_{\rm cut-5} = 
 \frac{ 4\,\,{\cal I}^{\rm CHY}_{\rm 4-Triangle}(1|2,3| 4 | [a_1,b_1],[a_2,b_2])}{(k_{a_1}+k_{b_1})^2\,\,(k_{a_2}+k_{b_2})^2}  + \ldots ,
\end{align*} 
and therefore the spurious pole on the support  $ \delta^{(D)} ( k_{a_2}+k_{a_1}) \,\,\delta^{(D)} (k_{b_2}+k_{b_1} ) $ is eliminated.

Unlike the bubble and triangle examples, the term ${\cal I}^{\rm CHY}_{\rm 4-Triangle}(1|2,3| 4 | [a_1,b_1],[a_2,b_2])$ does not vanish when $k_{a_2}=-k_{a_1}$ and $k_{b_2}=-k_{b_1}$,  and its result is given by 
\begin{align*}
&{\cal I}^{\rm CHY}_{\rm 4-Triangle}(1|2,3| 4 | [a_1,b_1],[a_2,b_2])  =  \frac{1}{k_{23}} \times\\
&\left\{
\frac{B ( \{ 2,3 \}, \{ 4 \} , \{ 1,a_1,b_1 \} ,\{ a_2,b_2 \})   }{(k_2+k_3+k_4)\cdot (k_{a_2}+k_{b_2})+k_{234}}
 + \frac{B ( \{ 2,3 \}, \{ 4 \} , \{ 1,a_2,b_2 \} ,\{ a_1,b_1 \})   }{(k_2+k_3+k_4)\cdot (k_{a_1}+k_{b_1})+k_{234}} \right. \\
& \left.  +
\frac{B ( \{ 2,3 \}, \{ 1 \} , \{ 4,a_1,b_1 \} ,\{ a_2,b_2 \})   }{k_4\cdot (k_{a_1}+k_{b_1})}
 +  \frac{B ( \{ 2,3 \}, \{ 1 \} , \{ 4,a_2,b_2 \} ,\{ a_1,b_1 \})   }{k_4\cdot (k_{a_2}+k_{b_2})}
 \right\}.
\end{align*}

Finally, after removing the spurious pole and on the support, $ \delta^{(D)} ( k_{a_2}+k_{a_1}) \,\,\delta^{(D)} (k_{b_2}+k_{b_1} ) $, one can  show
\begin{align}
\sum_{i=1}^8 {\bf I}^{\rm CHY}_{{\rm cut - }i}&=\frac{ 2^5}{s_{a_1b_1}}\times\frac{1}{s_{23}\,s_{a_1b_1}\,(k_{a_1}+k_{b_1}+k_{1})^2\, (k_{a_1}+k_{b_1}+k_{1}+k_{\{2,3 \}})^2}+{\rm Permutations}\{ 1,\{2,3\},4  \}
\nonumber
\\
&
=\frac{2^5}{s_{a_1b_1}} \times 
\sum_{p \in P_3}\frac{1}{(k_2+k_3)^2\,(k_{a_1}+k_{b_1})^2\,(k_{a_1}+k_{b_1}+k_{p_1})^2\, (k_{a_1}+k_{b_1}+k_{p_1}+k_{p_2})^2}\nonumber
\end{align}
where $P_3$ is defined as 
\begin{equation}
P_3:={\rm permutations} \,\{ \b_1,\b_2,\b_3\},\qquad {\rm with}~~\b_1:=1\,,\b_2:=\{2,3\}\,,\b_3:=4,
\end{equation}
for example, $k_{\b_2} = k_{\{2,3\}}=k_2+k_3$.

\begin{figure}[!h]
  \centering
         \includegraphics[width=1.6in]{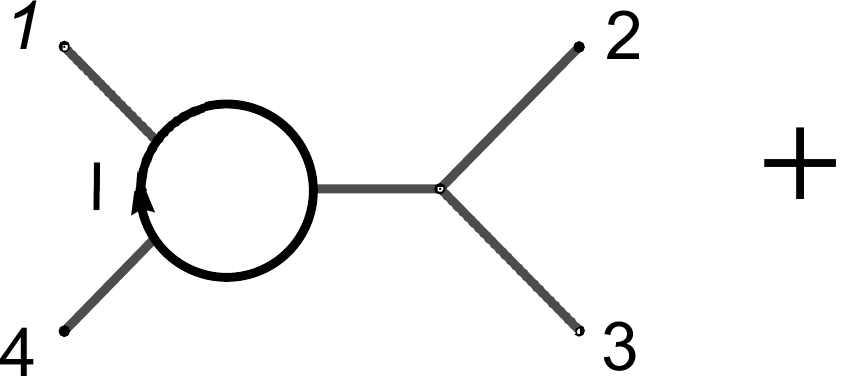}\quad
        \raisebox{2.3\height}{\includegraphics[scale=0.55]{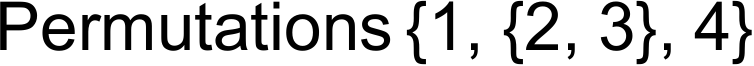}}
  \caption{Symmetrized Feynman diagram of the Triangle with four particles.}\label{fey-4p}
\end{figure}

Therefore, by integrating $\int\,d\Omega$ one obtains
\begin{align}
\mathfrak{I}^{\rm 3-gon-CHY}_{\rm sym} &=\int d\Omega\times \frac{(k_{a_1}+k_{b_1})^2}{2^5}\times 
\left(\sum_{i=1}^8 {\bf I}^{\rm CHY}_{{\rm cut - }i}\right)\nonumber\\
&=\frac{1}{s_{23}\,\ell^2\, (\ell+k_{1})^2\, (\ell+k_{1}+k_{\{ 2,3\}})^2}+{\rm Permutations\,\{1,\{2,3\},4 \}}
\nonumber\\
&=\frac{1}{s_{23}}\sum_{p\in P_3}\frac{1}{\ell^2\,(\ell+k_{p_1})^2\, (\ell+k_{p_1}+k_{p_2})^2},
\end{align}
which is the Feynman integrand of the sum of diagrams given in figure \ref{fey-4p}, as it was  expected.
%%%%%%%%%%%%%%%%%%%%%%%%%%%%%
\noindent
\\

%%%%%%%%%%%%%%%%%%%
\section{Discussions}\label{conclusion}
%%%%%%%%%%%%%%%%%%%

In this paper we have presented a new proposal in order to obtain the quadratic Feynman integrand at one-loop  from the CHY approach.  We have focused our research just to $\Phi^3$ theory, which is the simplest case, nevertheless,  we have already obtained some progress by extending our ideas to other theories \cite{wp}.

Note that our proposal is totally different to ones knew so far  \cite{Geyer:2015bja,Geyer:2015jch,Adamo:2013tsa,Baadsgaard:2015hia,He:2015yua,Cachazo:2015aol,Chen:2016fgi,Chen:2017edo,Cardona:2016bpi,Cardona:2016wcr}. Basically,  in all of these papers, the CHY approach of $n-$particles at one-loop is given by a contour integral over the moduli space of $(n+2)-$punctured spheres  with two off-shell particles in the forward limit (the internal loop momentum), and the final answer is obtained in the ${\cal Q}-$cut language \cite{Baadsgaard:2015twa}, such as in section \ref{sectionone}. In contrast, our formula is given by an integral over the moduli space of $(n+4)-$punctured spheres and the all particles are massless (section \ref{sectwo}). This is the first proposal where the CHY approach is able to obtained cleanly a quadratic Feynman integrand at loop-level, i.e. without to use the partial fraction identity\footnote{In other words, our result is not given in the ${\cal Q}-$cut representation.} nor shifting the internal loop momentum\footnote{Let us remind that  shifting on the loop moment  can lead to strong assumptions and consequences  both the integration contour over the internal loop  and its regulator \cite{Baadsgaard:2015twa}.}.

Schematically, one can say that our proposal is based on the idea that each internal loop puncture is interpreted as a couple of massless particles. This is the reason why a CHY-graph at one-loop, such as one drew in figure \ref{CHY-pgon-sym},  always appears in the computation, which is multiplied by a spurious pole that becomes singular on the support, $\delta^{(D)}(k_{a_2}+k_{a_1})\, \delta^{(D)}(k_{b_2}+k_{b_1})$, such as it was shown in the examples in section \ref{examples}.  However, our really motivation comes from the CHY prescription at two-loop developed by the author {\it et al} in \cite{Gomez:2016cqb}. 
%Note that form the formula proposed in \eqref{prescription}
%%%%%%%%%%%%%%%%%%%
\begin{figure}[!h]
%   Requires \usepackage{graphicx}
\centering
\raisebox{1.1\height}{\includegraphics[scale=.15]{Sum_Sp-eps-converted-to.pdf}}
 \includegraphics[scale=0.6]{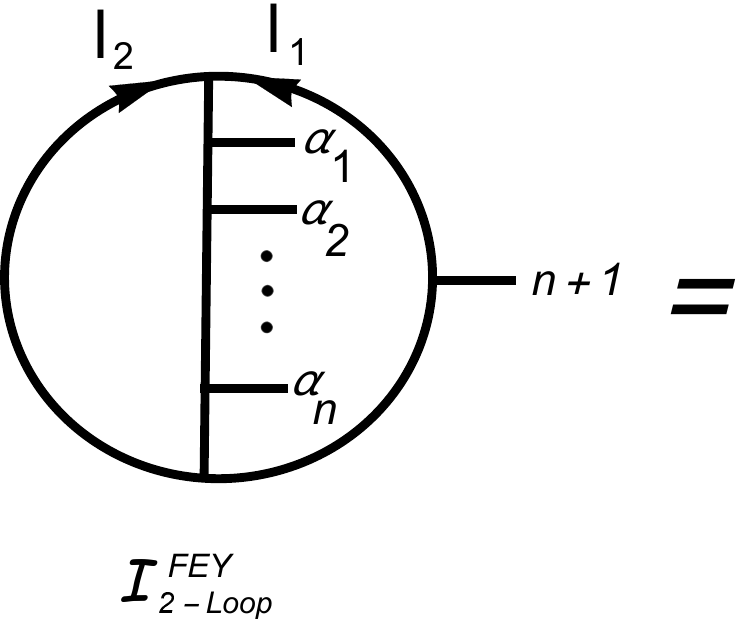}
 \raisebox{1.1\height}{\includegraphics[scale=.15]{Sum_Sp-eps-converted-to.pdf}}
 \includegraphics[scale=0.6]{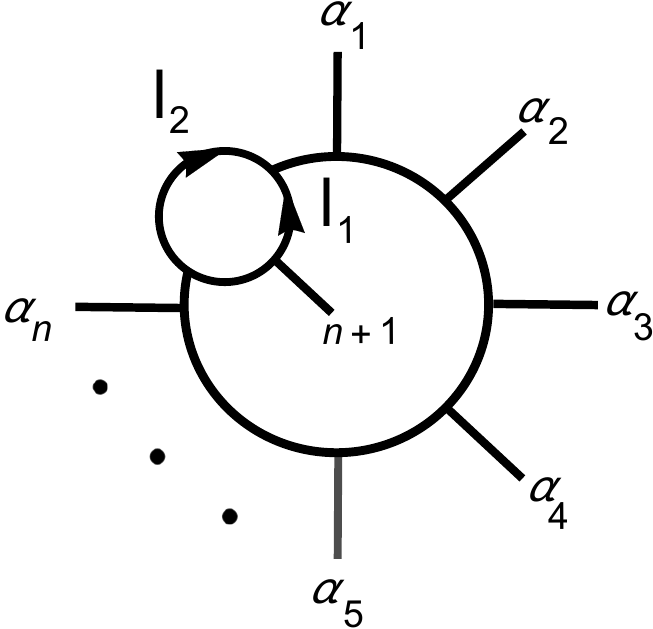}
  \caption{$\Phi^3$ Feynman diagram at two-loop.}\label{fey_2_loop}
\end{figure}
%%%%%%%%%%%%%%%%%%%%%%%%%%%%%
\noindent
\\

\begin{figure}[!h]
%   Requires \usepackage{graphicx}
\centering
 \includegraphics[scale=0.48]{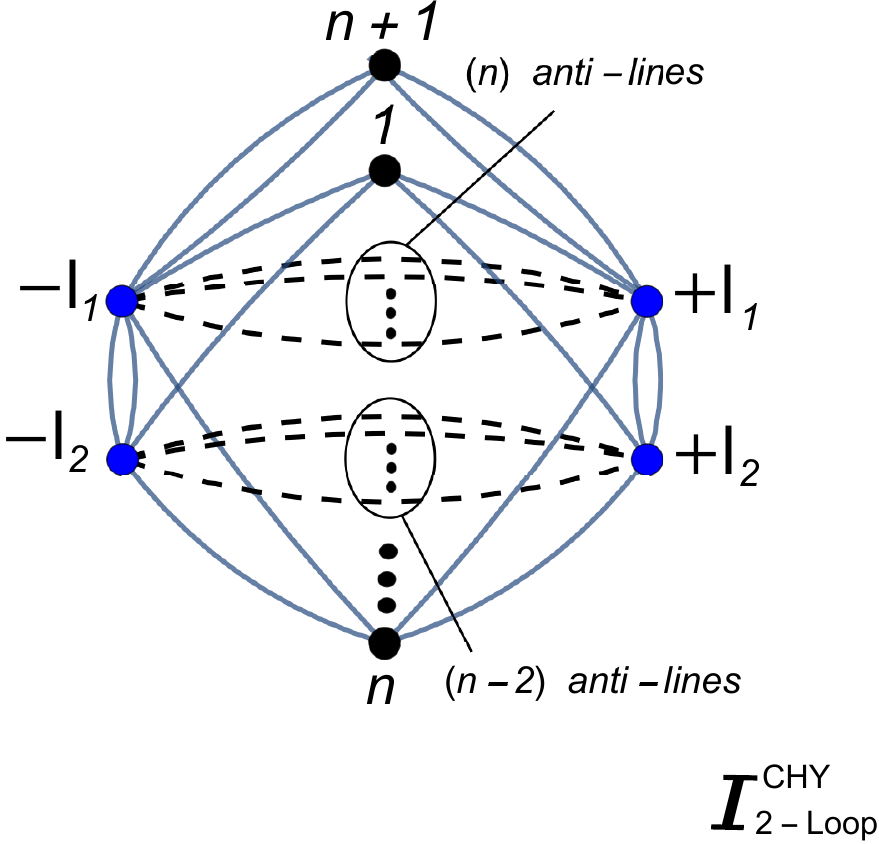}
 \raisebox{0.2\height}{\includegraphics[scale=.48]{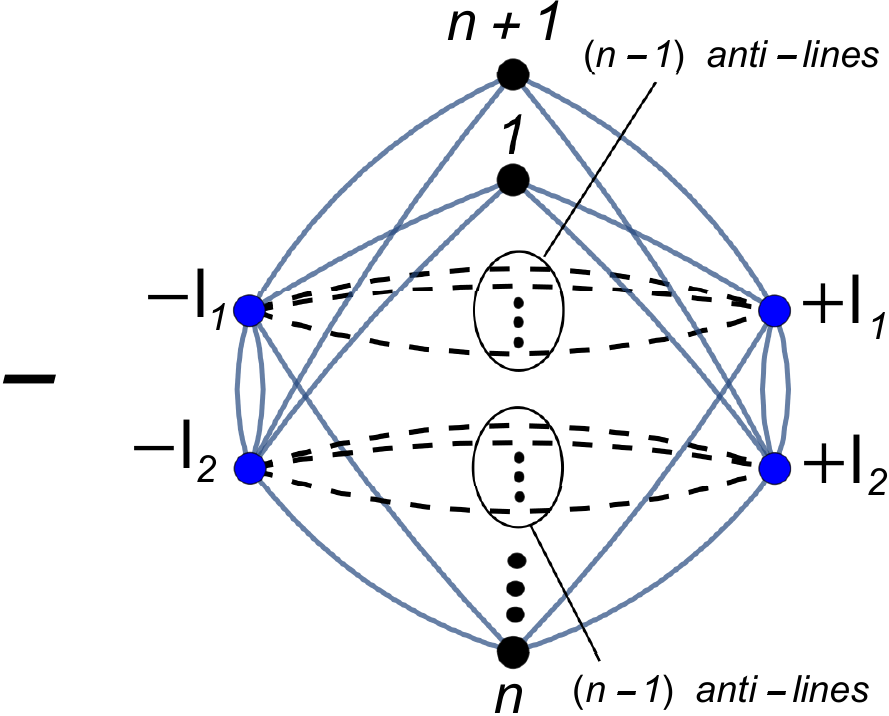}}
  \caption{CHY-graph representation for the ${\bf I}^{\rm CHY}_{\rm 2-Loop}$ integrand in \eqref{ICHY2L}.}\label{chy_2_loop}
\end{figure}

To understand the idea, let us consider the $\Phi^3$ Feynman diagram at two-loop given in figure \ref{fey_2_loop}, which has as integrand
\begin{equation}\label{fey_last}
{\cal I}_{\rm 2-loop}^{\rm FEY}= \frac{1}{\ell_1^2 \,\ell_2^2\, (\ell_1-k_{n+1})^2}\sum_{\a\in S_n}\frac{1}{(\ell_1+\ell_2)^2\, (\ell_1+\ell_2+k_{\a_1})^2\,\cdots (\ell_1+\ell_2+k_{\a_1}+\cdot+ k_{\a_n})^2}.
\nonumber
\end{equation}
Following the rules and building blocks  given by the author {\it et al} in \cite{Gomez:2016cqb}, the corresponding CHY-integral that reproduces  the ${\cal I}_{\rm 2-loop}^{\rm FEY}$ integrand, after partial fraction identity and shifting the loop momentum, is given by the expression
\begin{align}\label{ICHY2L}
{\cal I}_{\rm 2-Loop}^{\rm CHY}&=\frac{1}{\ell_1^2 \, \ell_2^2 }\int\, d\mu^{\rm 2-Loop}\, {\bf I}^{\rm CHY}_{\rm 2-Loop} ,\\
{\bf I}_{\rm 2-Loop}^{\rm CHY}&=\frac{1}{(\ell_1^+,\ell_2^+,\ell_2^-,\ell_2^-)^2} \,\,\left(\o_{1:2}^{\ell_1^+:\ell_1^-}\cdots \o_{n-1:n}^{\ell_1^+:\ell_1^-}\,\o_{n:1}^{\ell_1^+:\ell_1^-}\right)\times \left(\o_{1:n}^{\ell_2^+:\ell_2^-}\,\o_{n:n-1}^{\ell_2^+:\ell_2^-}\cdots \o_{2:1}^{\ell_2^+:\ell_2^-}\right) \nonumber\\
&\qquad \qquad \qquad \quad \quad
 \times \o_{n+1:n+1}^{\ell_1^+:\ell_1^-}  \left( \o_{n+1:n+1}^{\ell_1^+:\ell_1^-}  - \o_{n+1:n+1}^{\ell_2^+:\ell_2^-} \right),
 \nonumber
\end{align}
where the measure $d\mu^{\rm 2-loop}$ is defined in \cite{Geyer:2016wjx,Gomez:2016cqb} and 
the ${\bf I}_{\rm 2-Loop}^{\rm CHY}$ integrand is represented graphically in figure \ref{chy_2_loop}.
%%%%%%%%%%%%%%%%%%%
%%%%%%%%%%%%%%%%%%%%%%%%%%%%%
\noindent
\\
In \cite{Gomez:2016cqb},  it was argued that after carrying out the \eqref{ICHY2L} integral,  then ${\cal I}_{\rm 2-Loop}^{\rm CHY}$ becomes
\begin{align}
{\cal I}_{\rm 2-Loop}^{\rm CHY}&=\frac{2^n}{\ell_1^2 \,\ell_2^2\,( -\ell_1\cdot k_{n+1})}\sum_{\a\in S_n}\frac{1}{(\ell_1+\ell_2)^2\, (\ell_1+\ell_2+k_{\a_1})^2\,\cdots (\ell_1+\ell_2+k_{\a_1}+\cdots + k_{\a_n})^2}\nonumber \\
&\quad
+
\frac{2^n}{\ell_1^2 \,\ell_2^2\,( \ell_1\cdot k_{n+1})}\sum_{\a\in S_n}\frac{1}{(\ell_1+\ell_2)^2\, (\ell_1+\ell_2-k_{\a_1})^2\,\cdots (\ell_1+\ell_2-k_{\a_1}-\cdots - k_{\a_n})^2}\nonumber.
\end{align}
It is straightforward to check that using the partial fraction identity over the factor, $\frac{1}{\ell_1^2 (\ell_1-k_{n+1})^2}$, in ${\cal I}_{\rm 2-Loop}^{\rm FEY}$  and by shifting the $\ell_1^\mu$ loop momentum,  one obtains ${\cal I}_{\rm 2-Loop}^{\rm CHY}$.

\begin{figure}[!h]
  \centering
                    \includegraphics[scale=0.48]{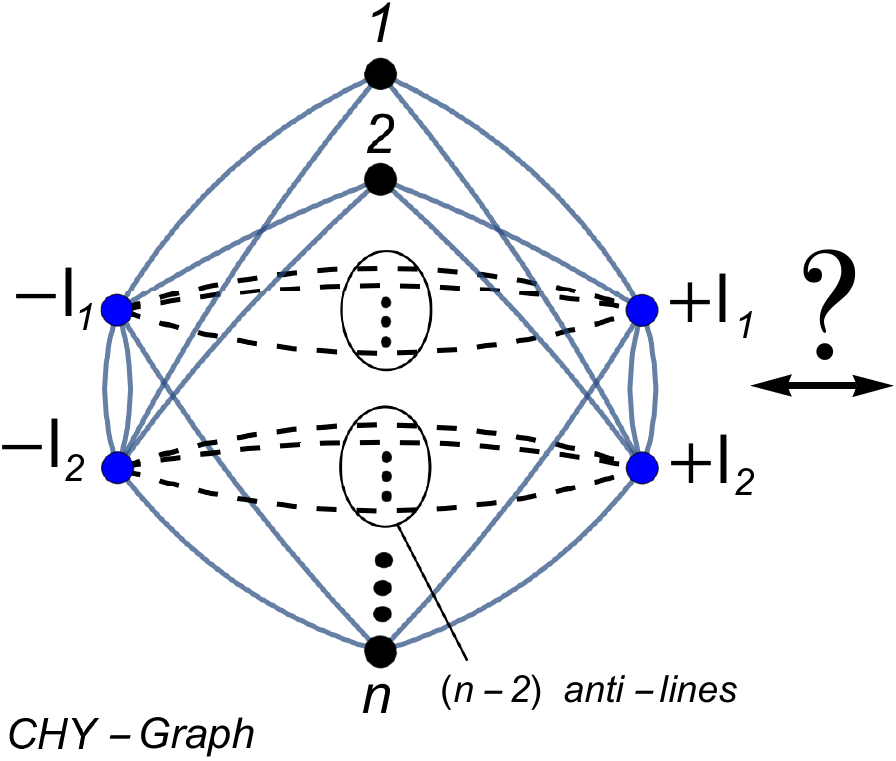} 
   \raisebox{1.1\height}{\includegraphics[scale=.15]{Sum_Sp-eps-converted-to.pdf}}
        \includegraphics[scale=0.6]{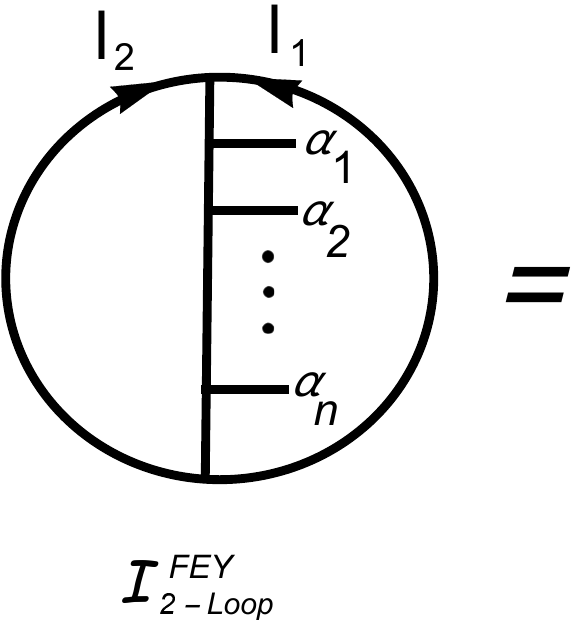}     
        \raisebox{1.1\height}{\includegraphics[scale=.15]{Sum_Sp-eps-converted-to.pdf}}                                         
         \includegraphics[scale=0.6]{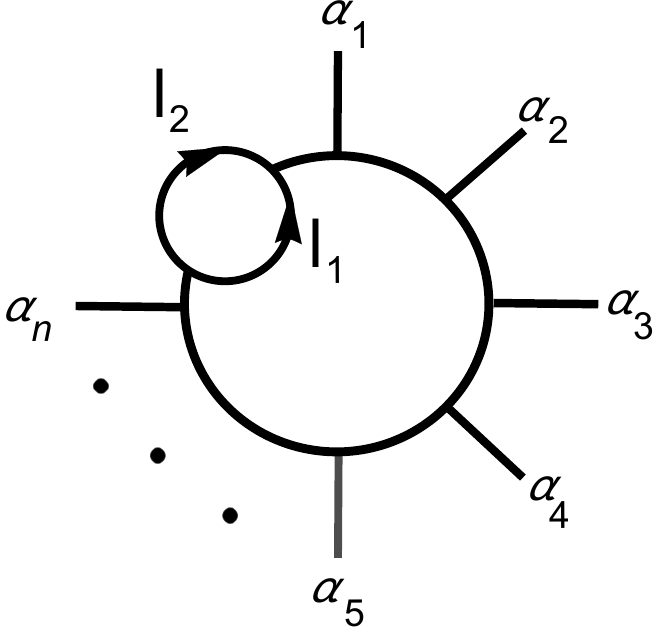}   
  \caption{CHY-graph that naively would be able to produce just quadratic Feynman propagators.}\label{qfp}
\end{figure}

\begin{figure}[!h]
%   Requires \usepackage{graphicx}
\centering
 \includegraphics[scale=0.45]{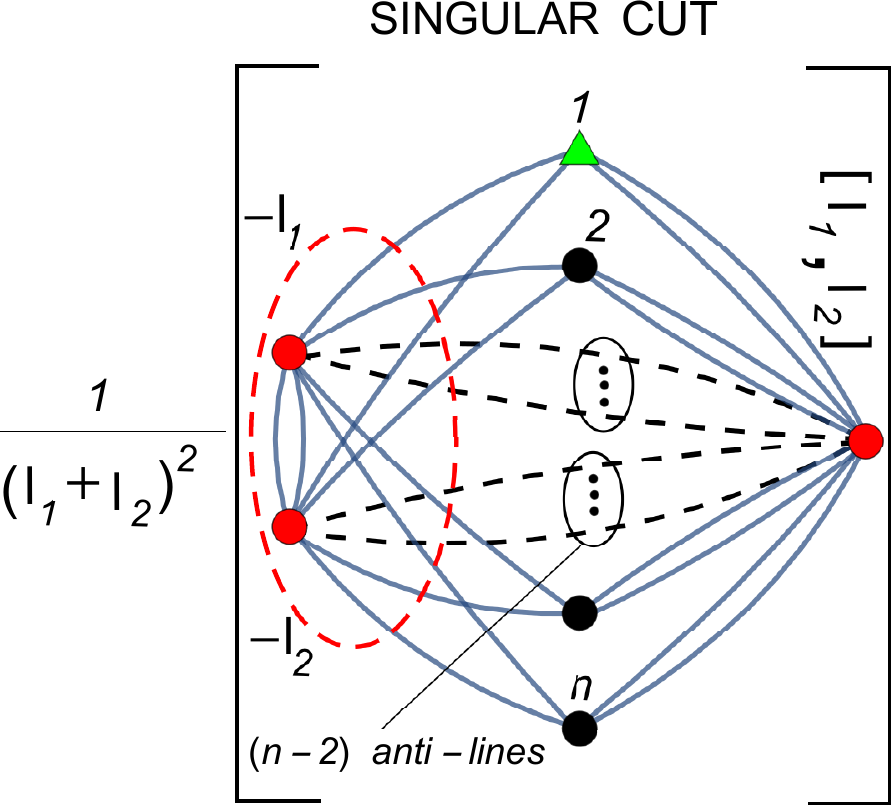}
  \caption{Singular cut (Tacnode singularity).}\label{singular_cut}
\end{figure}

Note that the CHY-integral in \eqref{ICHY2L} is able to produce some quadratic Feynman propagators, to be more precise, the propagators that are on the middle line  in the Feynman diagram drawn in figure \ref{chy_2_loop}.  So, in order to obtain just quadratic Feynman propagators, one can naively think in the CHY-graph given in figure \ref{qfp}. 
%%%%%%%%%%%%%%%%%%%
%%%%%%%%%%%%%%%%%%%%%%%%%%%%%
\noindent
\\
Nevertheless, an equivalence among the CHY-graph (left side) and the Feynman diagram (right side) in figure \ref{qfp} is not established, because the CHY-graph is singular.  From the $\L-$algorithm point of view this means there are divergent cuts, for example the one given in figure \ref{singular_cut}.  This singular cut is obtained by cutting $(\ell_1^+,\ell_2^+)$ and latter $(\ell_1^-,\ell_2^-)$,  which generates a CHY-graph at one-loop as in figure \ref{CHY-pgon-sym}, times the infinite propagator\footnote{This propagator becomes infinite by the momentum conservation condition, $\sum_{i=1}^nk_i=0$.},  $\frac{1}{(\ell_1+\ell_2)\cdot (k_1+\ldots + k_n) +k_{1\cdots n}}$. 
%%%%%%%%%%%%%%%%%%%
%%%%%%%%%%%%%%%%%%%%%%%%%%%%%
\noindent
\\
Notice that each CHY-graph in figure \ref{chy_2_loop} generates also these kind of singular cuts, but they are canceled out by the linear combination between  the graphs, which does not happen with the one in figure \ref{qfp}.\\
The geometrical meaning of this singularity can be studied from a hyper-elliptic curve of genus two. Let us consider the complex curve
\begin{equation}
y^2= z (z-1)(z-\l_1)(z-\l_2)(z-\l_3),
\end{equation}
where $ (\l_1, \l_2, \l_3) $ parameterizes the Moduli space (${\cal M}_{2}$) of this curve.  If $\l_1=0$ and $\l_2=1$ then the curve is degenerated to a sphere with four punctures and the parameter, $\l_3$, can be used to perform the $\L-$algorithm. This singularity is known as {\it node singularity}, which is equivalent to pinching two $A-$cycles, and it gives arisen to CHY-graphs as ones drawn in figure \ref{chy_2_loop} and \ref{qfp}.  Other type of singularity is, for example, when $\l_1=\l_2=\l_3=0$, which is known as {\it tacnode singularity}. These singularities  generate CHY-graphs as one given in figure \ref{singular_cut}, which has a propagator trivially infinite  by momentum conservation, for instance $\frac{1}{(\ell_1+\ell_2)\cdot (k_1+\ldots + k_n) +k_{1\cdots n}}$. So, to cancel out the tacnode singularities we consider a linear combination of graphs, as in figure \ref{chy_2_loop} \cite{Gomez:2016cqb}.  But, clearly the CHY-graph in figure \ref{qfp} is not able to do that.  Finally, the other types of singularities do not appear in the computation, so we do not consider them here.

\begin{figure}[!h]
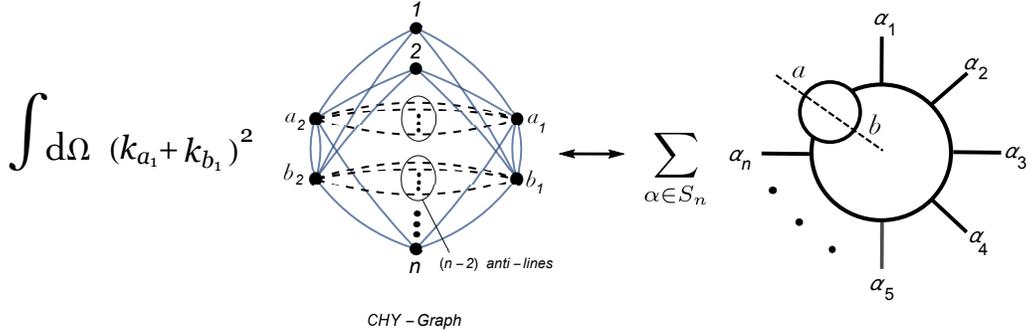

%   Requires \usepackage{graphicx}
\centering
\raisebox{2.0\height}{\includegraphics[scale=.5]{int_dO-eps-converted-to.pdf}}\quad
 \includegraphics[scale=0.5]{chy_Ucut-eps-converted-to.pdf}
\raisebox{1.4\height}{\includegraphics[scale=.15]{Sum_Sp-eps-converted-to.pdf}}
 \raisebox{0.1\height}{\includegraphics[scale=0.6]{FEY_Ucut-eps-converted-to.pdf}}
  \caption{Feynman diagram meaning for the new CHY proposal .}\label{Unicut}
\end{figure}

\begin{figure}[!h]
%   Requires \usepackage{graphicx}
\centering
 \includegraphics[scale=0.5]{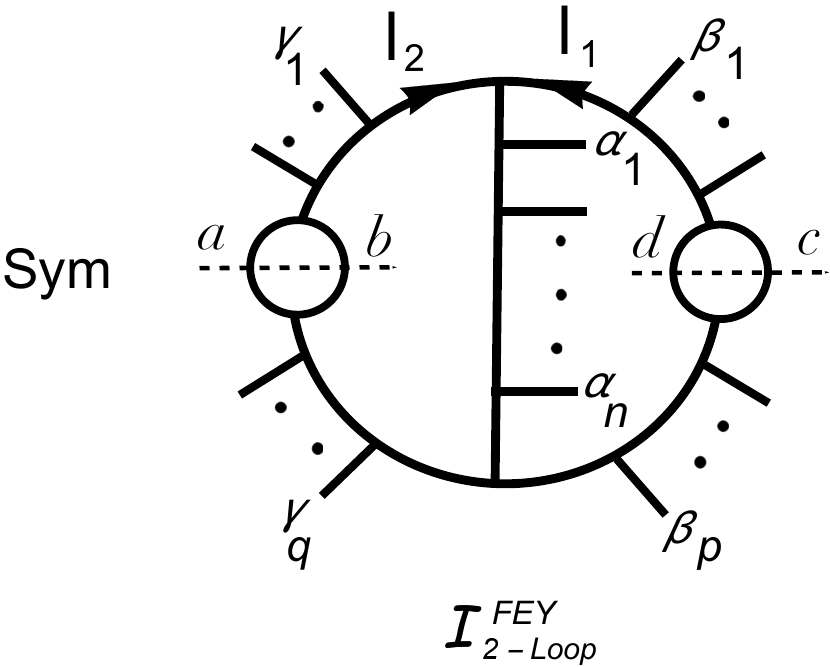}
  \caption{Representation of the process to obtain quadratic Feynman propagators at two-loop.
  ${\rm Sym}$ means sum over all permutations for each set of  external legs.}\label{four_loop}
\end{figure}

Our proposal is supported and motivated on the above ideas at two-loop \cite{Gomez:2016cqb}. Since we just wish obtained quadratic Feynman propagators, then it is natural to think over the CHY-graph in figure \ref{qfp}, again. But, so as to avoid the tacnode singularities, we consider all particles are differents, namely their momenta are generic. Naively, one could try to apply this trick on the graph in figure \ref{qfp}, and then making the forward limit, however, it could break the ${\rm PSL}(2,\mathbb{C})$ invariance of the scattering equations at two-loop.  In order to solve this drawback we regard the all particles are massless and therefore the scattering equations are the original ones given in \cite{Cachazo:2013hca,Cachazo:2013iaa,Cachazo:2013gna}. Now, to obtain a loop we should come back to try a forward limit, but, it generates a propagator trivially infinite multiplied by the tacnode singularity contribution, as it was seen previously. As we have shown in section \ref{examples},  this infinite propagator is in fact a fake infinity, which can be removed using the momentum conservation condition before making the forward limit.  Therefore, we now are able to  perform, transparently, the forward limit and, unlike with the two-loop case, the tacnode singularity is going to contribute to the computation. Finally, in order to obtain an internal loop we need an off-shell momentum. So, since the on-shell momenta related  with the  punctures generated by the node singularities, in figure \ref{Unicut} they are $(a_1,b_1)$ and $(a_2,b_2)$,  are always together as a couple, then we may identify this couple with an off-shell loop momentum, i.e. we are going to obtain an amplitude at one-loop.
%%%%%%%%%%%%%%%%%%%
%%%%%%%%%%%%%%%%%%%%%%%%%%%%%
\noindent
\\
In conclusion, we have developed a technique to obtain quadratic Feynman propagators at one-loop  from  a Riemann surfaces of genus two  by performing an unitary cut, such as it is shown in figure \ref{Unicut}, where $d\Omega$ is the meausre defined in \eqref{dOmega}. Therefore, if we wish to obtain the exact quadratic Feynman integrand at two-loop  given in  \eqref{fey_last}, we should perform an unitary cut  on a Riemann surface of genus four, as we schematized in figure \ref{four_loop}. Roughly speaking, this means that each off-shell puncture in figure \ref{chy_2_loop} should be splitting in two massless punctures \cite{wp}.
%Note that form the formula proposed in \eqref{prescription}
%%%%%%%%%%%%%%%%%%%
%%%%%%%%%%%%%%%%%%%%%%%%%%%%%
\noindent
\\

On the other hand, since the most of relationship among scattering amplitudes have been deeply studied  at tree-level, such as the Bern-Carrasco-Johansson (BCJ) relations, the Kawai--Lewellen--Tye (KLT) kernel,  monodromy relations or the soft limit behavior  \cite{Bern:2008qj,Kawai:1985xq,He:2016mzd,Tourkine:2016bak,Bjerrum-Bohr:2016axv,Bjerrum-Bohr:2016juj,Huang:2017ydz,Teng:2017tbo,He:2016vfi,Schwab:2014xua,Cachazo:2015ksa,Kalousios:2014uva,Afkhami-Jeddi:2014fia,Cachazo:2016njl,Stieberger:2016lng,
Hohenegger:2017kqy},  then following the ideas presented in this paper, where we have developed a technique to write scattering amplitudes of $n-$particles at one-loop as amplitudes of $(n+4)-$particles at tree-level,  we are confident that it is possible to apply the whole knowledge obtained at tree-level to find new relationships at loop-level.

\acknowledgments

We thank F. Cachazo, S. Mizera and G. Zhang for many useful discussions and comments. We are very grateful to the Perimeter Institute for hospitality. We would like to thank to P. Damgaard for many useful comments.
This research is supported by USC grant DGI-COCEIN-No 935-621115-N22.

%\newpage

\appendix

%%%%%%% 5pt Details %%%%%%%
%%%%%%%%%%%%%%%%%%%%%%%%%%%

\bibliographystyle{JHEP}
\bibliography{mybib}
\end{document}